# Dissipative quadratic soliton mode-locked optical parametric oscillator

Jonathan Musgrave[1], Mingming Nie[1, *, †], and Shu-Wei Huang[1, *]

[1]*Department of Electrical, Computer and Energy Engineering, University of Colorado Boulder, Colorado 80309, USA*

[*]*Corresponding authors: mingming.nie@uestc.edu.cn, shuwei.huang@colorado.edu*

[†]*Current Address: Key Laboratory of Optical Fiber Sensing and Communications (Education Ministry of China), University of Electronic Science and Technology of China, Chengdu 611731, China*

**Abstract:** Femtosecond lasers enable ultrafast spectroscopy and imaging, but their wavelength coverage is limited by the emission bands of available laser materials. Optical parametric oscillators extend this coverage through nonlinear frequency conversion, yet their femtosecond operation typically requires synchronized pumping by an external pulsed laser. Here we show that continuous-wave pumping can spontaneously generate femtosecond pulses in an optical parametric oscillator through passive dissipative quadratic soliton mode-locking. These dissipative quadratic solitons are stable pulses sustained by a balance between nonlinear frequency conversion, dispersive spreading, gain and loss. Repeated conversion between pump and generated light in a cavity resonant at both wavelengths creates an effective Kerr nonlinearity, an intensity-dependent phase response, over a thousand times stronger than the material's intrinsic Kerr nonlinearity. Its magnitude and sign are adjustable through the pump's frequency offset from a cavity resonance, enabling soliton formation with either sign of dispersion. Experimentally, 600 mW of continuous-wave pumping yields pulses at 1572 nm and 786 nm with durations of 336 fs and 447 fs, respectively, peak powers up to 150 W and 5% pump-to-soliton conversion efficiency. These results establish a simpler architecture for femtosecond optical parametric oscillators and introduce tunable nonlinearity as a design principle for extending soliton generation across photonic platforms and spectral regions.

## Introduction

Femtosecond mode-locked lasers (MLL), most notably Ti:sapphire and rare-earth-doped fiber lasers, have transformed ultrafast science, yet their spectral reach is limited by the finite emission bandwidths of available laser materials. Optical parametric oscillators (OPOs) overcome these constraints by exploiting second-order nonlinear interactions to deliver widely tunable, phase-coherent radiation across spectral regions inaccessible to direct laser emission, including the mid-infrared and ultraviolet [1–3]. Femtosecond OPOs therefore support ultrafast spectroscopy and nonlinear microscopy at wavelengths resonant with molecular vibrations and electronic transitions beyond conventional laser coverage [4–10]. The deterministic phase relationships of synchronously pumped OPOs also have enabled emerging non-von Neumann computing paradigms, notably coherent Ising machines for NP-hard combinatorial optimization [11–16]. By generating squeezed states and entangled photon pairs, OPOs also serve as key platforms for quantum-enhanced sensing and high-dimensional quantum information science [17–21].

Femtosecond OPOs have traditionally used a dispersion-free cavity synchronously pumped by an external femtosecond MLL [22–29]. Despite the unmatched performance and spectral coverage, they rely on complex synchronous pumping schemes, expensive femtosecond MLLs, and precise control electronics, increasing system complexity, footprint and cost (Fig. 1a).

To address these issues, we exploit the physics of dissipative quadratic solitons (DQS) [30–40] to passively mode-lock a continuous-wave (CW) driven degenerate doubly resonant degenerate optical parametric oscillator (DR-DOPO). DQSs were recently shown to enable efficient pulse shaping in synchronously pumped OPOs [41,42], where strong pump-signal temporal walk-off drives substantial pulse compression from ~13-ps pump pulses to ~320-fs signal pulses. Here, we go beyond DQS pulse shaping and demonstrate a fundamentally different regime: spontaneous femtosecond DQS formation from a CW pump with minimized pump-signal temporal walk-off.

We present the operating principle and mean-field theory of the DQS mode-locked OPO and show that phase-matched intracavity cascaded quadratic nonlinearity (PICQN), existing only in a doubly resonant cavity with negligible pump-signal temporal walk-off, can efficiently generate effective Kerr nonlinearity (EKN) and effective two photon absorption (ETPA) that dominates the DQS mode-locking dynamics. We demonstrate this mechanism in a free-space periodically poled lithium niobate (PPLN) cavity with a greatly simplified pump system (Fig. 1b). With 600 mW of CW pump power, the signal DQS at 1572 nm has a spectral bandwidth of 1.21 THz, a pulse duration of 336 fs, and a peak power of 150 W, corresponding to a pump-to-DQS power conversion efficiency of 5%. The cascaded nature of PICQN manifests itself in the generation of a pump DQS at 786 nm, with a spectral bandwidth of 0.91 THz, a pulse duration of 447 fs, and a peak power of 80 W, producing a bright bichromatic frequency comb spanning the visible and near-infrared spectral ranges (Fig. 1b). Agreement between experiment and theory validates the theoretical framework and establishes a numerical model for designing and investigating the full family of DQS states. PICQN-enabled DQS broadens the design space for dissipative solitons, enabling flexible spectral tailoring and suggesting the possibility of universal dissipative soliton existence across photonic platforms and dispersion regimes.

## Results

### Operating Principle

Fig. 1b illustrates DQS mode-locking in the CW-pumped DR-DOPO. The coherent pump drives the cavity at finite pump phase detuning (bottom left). Cascaded quadratic conversion and back-conversion between pump and signal produce a nonlocal effective third-order nonlinearity (bottom middle). We previously demonstrated DQS in a singly resonant (SR) cavity, where phase mismatch alone drives the cascaded process [43]. In a DR cavity, phase mismatch and phase detuning jointly control the relative interaction phase, enabling PICQN and more efficient generation of an effective third-order nonlinearity tunable through pump phase detuning. We recently demonstrated DQS mode-locking of cavity-enhanced second-harmonic generation in the same DR cavity architecture under 1572-nm pumping [40]. Here, pumping at 786 nm changes the cavity injection boundary conditions, making the 1572-nm field parametrically driven rather than directly pumped and thereby producing distinct mode-locking dynamics. We precisely control pump phase detuning and, consequently, the PICQN using the auxiliary-pump scheme described in Methods. Supplementary Information Section I derives the governing equations of DQS mode-locked DR-DOPO dynamics, while Table S2 compares the EKN and ETPA contributions to the effective third-order nonlinearity in SR and DR cavities.

At sufficiently large pump phase detuning, EKN dominates /ETPA (Fig. 1c), yielding a mechanism analogous to dissipative Kerr soliton formation [44] (Fig. 1b, bottom right). Unlike conventional TPA, which induces loss through nonlinear electronic bandgap absorption, ETPA represents pump-signal energy transfer and only perturbs soliton formation. The true energy dissipation arises solely from the linear cavity losses. In this regime, the engineered EKN not only exceeds the intrinsic material Kerr nonlinearity (MKN) by over three orders of magnitude but is also continuously tunable in both magnitude and sign via pump phase detuning (Fig. 1c). This orders-of-magnitude enhancement of EKN enables free-space dissipative solitons without strong confinement in fibers or microresonators, allowing scalable power and energy. The in-situ tunability of EKN allows compensation of both normal and anomalous dispersion, shifting soliton design from dispersion to nonlinearity engineering. PICQN-enabled DQS thus broadens the design space for dissipative solitons, allowing flexible spectral control and suggesting universal soliton existence across platforms and dispersion regimes.

### Mean Field Theory

As shown in Supplementary Information Section I, the evolution of the intracavity signal DQS can be modeled with the parametrically driven Ginzburg-Landau equation [34]:

$$\partial_t \mathsf{A} = -(1+i\Delta_1)\mathsf{A} - i\beta_1\partial_\tau^2\mathsf{A} - \mathsf{A}^*\left(\mathsf{A}^2 \otimes \bar{J}(\tau)\right) + \rho\mathsf{A}^* \quad (1)$$

where t and $\tau$ are normalized slow (lab-frame) and fast (resonator) times, respectively. $\beta_1 = \mathrm{sgn}(k_1'')$ is the sign of the signal group velocity dispersion (GVD), $\rho$ is the parametric drive, and $\otimes$ denotes the convolution with the nonlocal nonlinear response

$$\bar{J}(\tau) = \mathrm{sinc}^2(\xi)\,\mathcal{F}\left\{\frac{\alpha(1+(\Delta_2/\alpha)^2)}{\alpha + i(\Delta_2 - d\Omega - \beta_2\Omega^2)}\right\}. \quad (2)$$

where $\mathcal{F}$ is the Fourier transform. $\Delta_{1,2}$ are signal and pump phase detuning, $\xi = \Delta kL/2$ is normalized phase-mismatch, $\alpha = \alpha_2/\alpha_1$ is the pump-to-signal cavity loss ratio, $\beta_2 = k_2''/|k_1''|$, $d$ is normalized group velocity mismatch (GVM), and $\Omega$ is normalized frequency offset from the signal carrier frequency $\omega_0$.

The pump is dynamically slaved to the signal:

$$\mathsf{B} = \left(-\mathsf{A}^2 \otimes \bar{J}(\tau) + \rho\right)e^{i\tan^{-1}(\Delta_2/\alpha)}\Big/\sqrt{\alpha(1+\Delta_2^2/\alpha^2)}. \quad (3)$$

Supplementary Information Section XIII provides a table summarizing all experimental parameter definitions and values. In particular, both pump and signal operate in the normal-dispersion regime, with $\beta_1 = 1$ and $\beta_2 = 3$. GVM and phase mismatch are both zero, namely $d = 0$ and $\xi = 0$. Pump and signal quality factors are comparable, yielding $\alpha = 1$. Finally, phase matching imposes $\Delta_2 = 2\Delta_1$ (Supplementary Information Section II).

The homogenous solutions of Equation (1) admit non-trivial solutions, $\mathsf{A}^{\pm}$ and a trivial solution $\mathsf{A}^0$. When $\Delta_2\Delta_1 < 1$, only $\mathsf{A}^+$ exists and bifurcates supercritically at the pitchfork bifurcation $\rho_a = \sqrt{1+\Delta_1^2}$ (black line, Fig. 2a). For $\Delta_2\Delta_1 > 1$, both $\mathsf{A}^+$ and $\mathsf{A}^-$ coexist and connect at the turning point $\mathrm{SN_t} = (\Delta_2 + \Delta_1)/\sqrt{1+\Delta_2}$ (blue line, Fig. 2a). Fig. 2b shows example bifurcation diagrams at $\Delta_1 = -0.7$ and $-5$, plotting $||\mathsf{A}||^2 = T^{-1}\int_{-T/2}^{T/2} |\mathsf{A}|^2 d\tau$ against $\rho$, where $T = 60$ is the simulation time window. Solid and dashed blue lines denote stable and unstable states. respectively.

These two bifurcations carve out three main regions in the $(\Delta_1, \rho)$-parameter space as shown in Fig. 2a.

I. Below threshold $\rho < \mathrm{SN_t}$, only $\mathsf{A}^0$ exists and is stable. A homogenous cavity solution with no generated signal for $(\Delta_1, \rho)$ = (-3,0.5) is shown.

II. For $\rho > \rho_a$, $\mathsf{A}^+$ and $\mathsf{A}^0$ coexist, with $\mathsf{A}^0$ unstable. A chaotic cavity solution for $(\Delta_1, \rho)$ =(-2,4) is shown.

III. For $\mathrm{SN_t} < \rho < \rho_a$, all three states coexist. A DQS solution for $(\Delta_1, \rho)$ = (-5,4.5) is shown.

In region III, $\mathsf{A}^0$ is stable whereas $\mathsf{A}^{\pm}$ need not be. Sufficient excitation, such as sweeping pump phase detuning into resonance, accesses stable periodic patterns (PP), or Turing patterns. These PPs emerge subcritically along $\mathsf{A}^+$ (dark yellow line, Fig. 2b) and coexist with $\mathsf{A}^0$, establishing Turing bistability. A complete stability analysis of this regime is provided in Supplementary Information Section

III. Here, patterned fronts between $A^0$ and PP solutions can spontaneously connect to form stable localized high-peak power DQS with no CW-background (see Fig. 2c), resulting in DQS mode-locked DR-DOPO.

We trace these states using Newton-Raphson continuation with pseudo-arclength prediction, starting from a weakly nonlinear solution near $(\Delta_1, \rho) = (-5, \rho_a)$. Details are outlined in Supplementary Information Section IV. At fixed $\rho$, Fig. 2c shows discrete energy levels from single DQSs (red) to seven-soliton molecules (bright yellow). Homoclinic snaking connects these states through saddle-node bifurcations (black dashed curves, inset). Successive Turing-pattern peaks nucleate at the bistability boundary, each adding a DQS, eventually filling the resonator with an equally spaced pulse train corresponding to PP on the trivial background.

Fig. 2d shows temporal and spectral profiles (left and right) of states i–iii: a single DQS and double- and triple-soliton molecules. Fig. 2e shows the zoom-in view of the single signal DQS intensity $|A|^2$ (red), the slaved pump DQS intensity $|B|^2$ (cyan-blue), and the pump spectrum. The instantaneous signal frequency, $\Omega(\tau) = \partial_\tau(\arg[A])$, reveals slight residual chirp (blue).

Our analysis shows that Turing bistability supports DQS mode-locking in both normal and anomalous dispersion regimes. Supplementary Information Section VI provides further details of DQS mode-locking in the opposite (anomalous) dispersion regime.

**Experimental Demonstration**

Guided by our stability analysis, we constructed a CW-pumped DR-DOPO with a free spectral range (FSR) of 380 MHz and a loaded quality factor of $2.70 \times 10^8$ (Fig. 1b). Pumping at 786 nm achieves zero GVM in type-I phase-matched PPLN. While type-I phase-matching reduces the effective nonlinear coefficient ($d_{eff}$) to 2.7 pm/V, birefringence enables independent tuning of phase matching and double resonance through precise control of the cavity length or PPLN crystal temperature [45]. The pump and signal GVDs are both normal at 378 $fs^2$/mm and 105 $fs^2$/mm, respectively, precluding parametrically driven dissipative Kerr soliton (DKS) formation. The cavity has no dispersion compensation element and thus the cavity dispersion is determined by the PPLN bulk material dispersion. Supplementary Information Section IX further examines how GVD compensation affects DQS spectral bandwidth and pulse duration. Further details of the experimental setup are provided in the Methods section.

The cavity parameters closely match those used in Fig. 2. At a pump power of 600 mW, corresponding to a normalized parametric pump amplitude of $\rho \approx 4.8$ at $\Delta_1 = -5.2$, we sweep the pump frequency from the red to the blue side of the cold-cavity resonance (Fig. 3). This sweep traverses the $(\Delta_1, \rho)$ parameter space horizontally (Fig. 2a). Simultaneously measured pump and signal optical spectra (Fig. 3a) and signal radio-frequency (RF) spectra (Fig. 3b) reveal the evolving cavity fields and signal coherence. Figures 3c,d show representative optical and corresponding RF spectra.

As the pump enters resonance from the red-detuned side, the intracavity power exceeds the OPO threshold, and PICQN generates primary comb lines (state I, $\Delta_1 = 4.5$), followed by secondary comb lines (state II, $\Delta_1 = 1.7$). These comb lines approach the degenerate signal frequency as phase anomalies associated with the nonlocal nonlinear response narrow (Supplementary Information Section I). On the blue-detuned side, the system enters region II of Fig. 2a. A strong CW peak at the degenerate signal wavelength represents the homogeneous solution, accompanied by a broad incoherent pedestal (state III, $\Delta_1 = -1.2$). Further blue detuning increases the pedestal's parametric gain, producing chaotic temporal evolution (state IV, $\Delta_1 = -3.3$). Correspondingly, the low-noise RF tone at the cavity repetition rate broadens (Fig. 3d, states i–iv). This incoherent state allows entry into the Turing bistability regime with a non-trivial and unstable cavity energy.

Further pump tuning spontaneously converts state IV into a stable single DQS (state V, Fig. 4a) through the locking of patterned fronts, consistent with the stability analysis in Fig. 2. The observed transient dynamics agree closely with numerical integration of the normalized mean-field equations (Supplementary Information Section V and Supplementary Video 1).

The signal and pump DQS spectra follow $\text{sech}^2$ profile (Fig. 4), with bandwidths of 1.21 THz and 0.91 THz and corresponding transform-limited pulse durations of 260 fs and 350 fs, respectively. DQS formation produces a sharp transition from the broad, incoherent RF spectrum to a low-noise tone with a signal-to-noise ratio (SNR) exceeding 60 dB (Fig. 4b), demonstrating high coherence and stability. The signal DQS single-sideband phase noise rolls off by 20-30 dB per decade at low offset frequencies and reaches a noise floor of −130 dB/Hz at a 100-kHz offset (Supplementary Information Section VII).

Frequency-resolved optical gating (FROG; Supplementary Information Section VIII) reconstructs the signal DQS pulse profile and instantaneous frequency (Fig. 4c, red and blue solid curves). Both agree well with theoretical predictions (corresponding dashed curves), with minor discrepancies attributed to higher-order dispersion omitted from the model. The signal pulse duration is 336 fs, with a linear chirp near its center, and external compression could approach the 260-fs transform limit. The pump-to-DQS power conversion efficiency is 5%, comparable to MLL efficiency, yielding a signal DQS peak power of 150 W with a 600-mW CW pump power. PICQN's cascaded nature also generates a pump DQS at 786 nm with a 0.91-THz bandwidth. Using the measured signal DQS time-bandwidth product of 0.406, we estimate the pump DQS pulse duration and peak power to be 447 fs and 80 W, respectively, resulting in a bright bichromatic frequency comb spanning the visible and near-infrared.

Repeated resonance sweeps access other predicted soliton states, including those labelled ii and iii in Fig. 2c, through stochastic evolution from the chaotic state IV. Figure 4d shows the final signal-spectrum trajectories from two sweeps that produce two- and three-soliton molecules, respectively, consistent with the stability analysis and numerical integration (Supplementary Information Section V and Supplementary Video 1).

Fig. 4e shows the dependence of DQS spectral bandwidth and pulse duration on the pump phase detuning (top) and pump power (bottom). These dependences resemble those of DKSs [46,47] but show the opposite trend to walk-off induced DQS pulse shaping [41], confirming that DQS mode-locking and pulse shaping arise from fundamentally different physical mechanisms.

## Discussion

Although EKN relaxes the need for dispersion engineering, reducing the net cavity dispersion remains an effective route to shorten DQS pulses. Simulations in Supplementary Information Section IX indicate that cavity-dispersion compensation could shorten the DQS pulse duration to approximately 30 fs. Regarding conversion efficiency, the analysis in Supplementary Information Section XI predicts pump-to-DQS conversion efficiencies well above 10% with appropriately chosen pump- and signal-cavity finesses. More broadly, the substantial design freedom of the DQS platform makes conversion-efficiency optimization a multivariate problem involving several interdependent parameters.

GVM is the primary perturbation to DQS mode-locking and the principal constraint on the spectral coverage (Supplementary Information Section X). Figure S12 maps wavelength ranges with tolerable GVM and shows the temperature tunability of zero-GVM wavelengths in common nonlinear crystals. Appropriate crystal and temperature selection enables coverage of much of the mid-infrared molecular-fingerprint region. Spectral coverage could be further extended by increasing cavity loss and/or detuning, at the expense of a higher pump threshold, or by compensating the net cavity GVM through engineered dielectric-mirror coatings [48], non-collinear geometries [49], material compensations [50], polarization multiplexing [51], or space multiplexing [52].

Parametrically driven DKS offer another route to passive DOPO mode-locking [53,54]. Although sharing certain similarities with DQS, it relies on conventional dispersion engineering to balance the intrinsic MKN. By contrast, DQS introduces a paradigm shift from dispersion to nonlinearity engineering for dissipative soliton formation. The engineered EKN exceeds the intrinsic MKN by more than three orders of magnitude and is continuously tunable in both magnitude and sign via pump phase detuning. This enhancement eliminates the need for strong mode confinement, enabling scalable power and energy. The in-situ tunability of EKN allows compensation of both normal and anomalous dispersion, suggesting the possibility of universal dissipative soliton existence.

The recent observation of topological soliton crystals in a CW-pumped, singly resonant DOPO [55] independently highlights the potential of quadratic soliton formation. However, the singly resonant architecture imposes a higher pump threshold, and the bright-dark soliton pair nature of these states leads to low peak power at the OPO wavelength, where the field is intrinsically a dark soliton. By contrast, our PICQN-enabled DQS mode-locked OPO generates bright solitons simultaneously at the pump and OPO wavelengths and enables access to a genuine single bichromatic soliton state beyond the soliton-crystal regime.

Our results establish a simple, flexible, and scalable route to femtosecond OPO operation without a synchronized mode-locked pump laser. PICQN-enabled DQS broadens soliton technologies across photonic platforms and dispersion regimes, offering a route to bichromatic femtosecond frequency combs at difficult-to-access wavelengths. Whereas the demonstrated free-space PPLN cavity supports power and energy scaling, thin-film PPLN photonic integrated circuits [56–59] offer a path toward scalable fabrication of compact, ultralow-threshold DQS comb sources spanning the visible to mid-infrared. Furthermore, extending DQS mode-locking to non-degenerate operation could enable real-time wavelength tuning at a fixed pump wavelength [37].

Beyond femtosecond OPOs, in situ nonlinearity engineering adds a degree of freedom to cavity nonlinear optics. For example, engineered EKN could compensate MKN and suppress modulation instability at high intracavity power, enhancing the entanglement-source efficiency for quantum photonic processing [60,61] and improving photonic sensor performance [62,63]. Reconfigurable EKN also opens opportunities for advanced photonic devices, including programmable nonlinear photonic circuits for artificial intelligence applications.

**Acknowledgments**
We thank TOPTICA Photonics for the loan of the BoosTA-pro 780 tapered amplifier.

**Funding Statement**
This work was supported by the Office of Naval Research (N00014-22-1-2224 to S.W.H.) and National Science Foundation (ECCS2048202 to S.W.H.). J.M. acknowledges the support of the Air Force Office of Scientific Research through the NDSEG fellowship.

**Author Contributions Statement**
M.N. and S.W.H. conceived the idea. J.M. and M.N. designed the experiment. J. M. performed the experiment, numerical simulation, and theoretical analysis. J.M., M.N., and S.W.H. conducted the data analysis and wrote the manuscript. S.W.H. led and supervised the whole project. All authors contributed to the discussion and revision of the manuscript.

**Competing Interests Statement**
J.M., M.N., and S.W.H. are the inventors of a provisional patent application, filed by the University of Colorado Boulder, about the dissipative quadratic soliton (DQS) frequency comb.

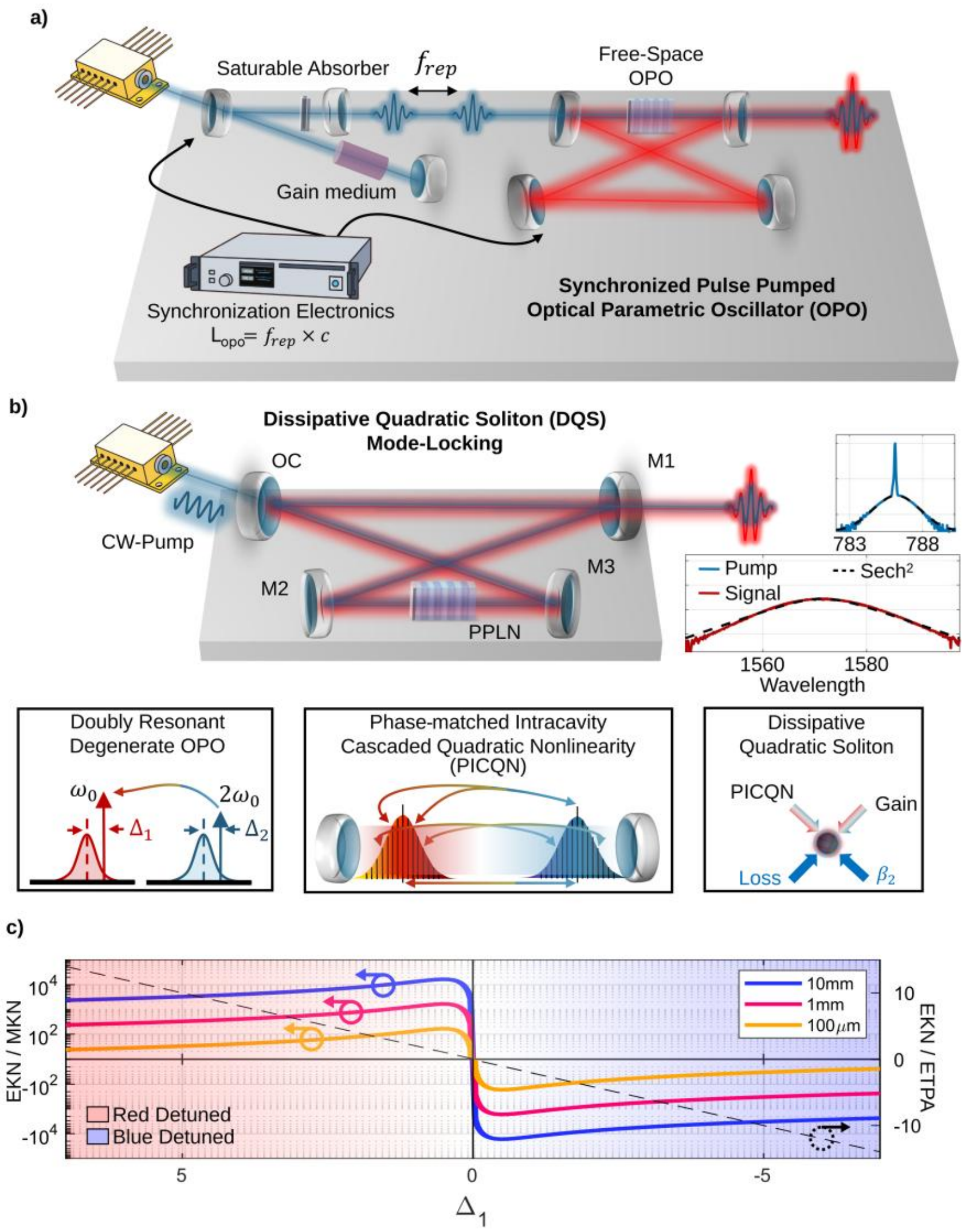

**Fig. 1 | DQS mode-locked degenerate optical parametric oscillator.** a) Conventional pulse-pumped OPOs require a dispersion-free resonant cavity, a femtosecond mode-locked pump laser and complex synchronization electronics. b) DQS mode-locking uses continuous-wave (CW) pumping with precisely controlled pump phase detuning (Methods), enabling a simple, flexible and scalable femtosecond OPO. The four-mirror free-space doubly resonant cavity contains a type-I phase-matched periodically poled lithium niobate (PPLN) crystal pumped by a *p*-polarized CW laser. The output coupler (OC) transmits 1% at 1572 and 786 nm; mirrors M1-M3 are highly reflective at both wavelengths. Bottom panels show coupled pump and signal detunings (left), effective nonlinearity and parametric gain from PICQN controlled by pump phase detuning (middle), and DQS formation balancing nonlinearity, dispersion, cavity loss and parametric gain (right). c) Ratios of effective Kerr nonlinearity (EKN) to material Kerr nonlinearity (MKN) for different crystal lengths (solid lines, left axis), and to effective two-photon absorption (ETPA) (dashed line, right axis), versus pump phase detuning.

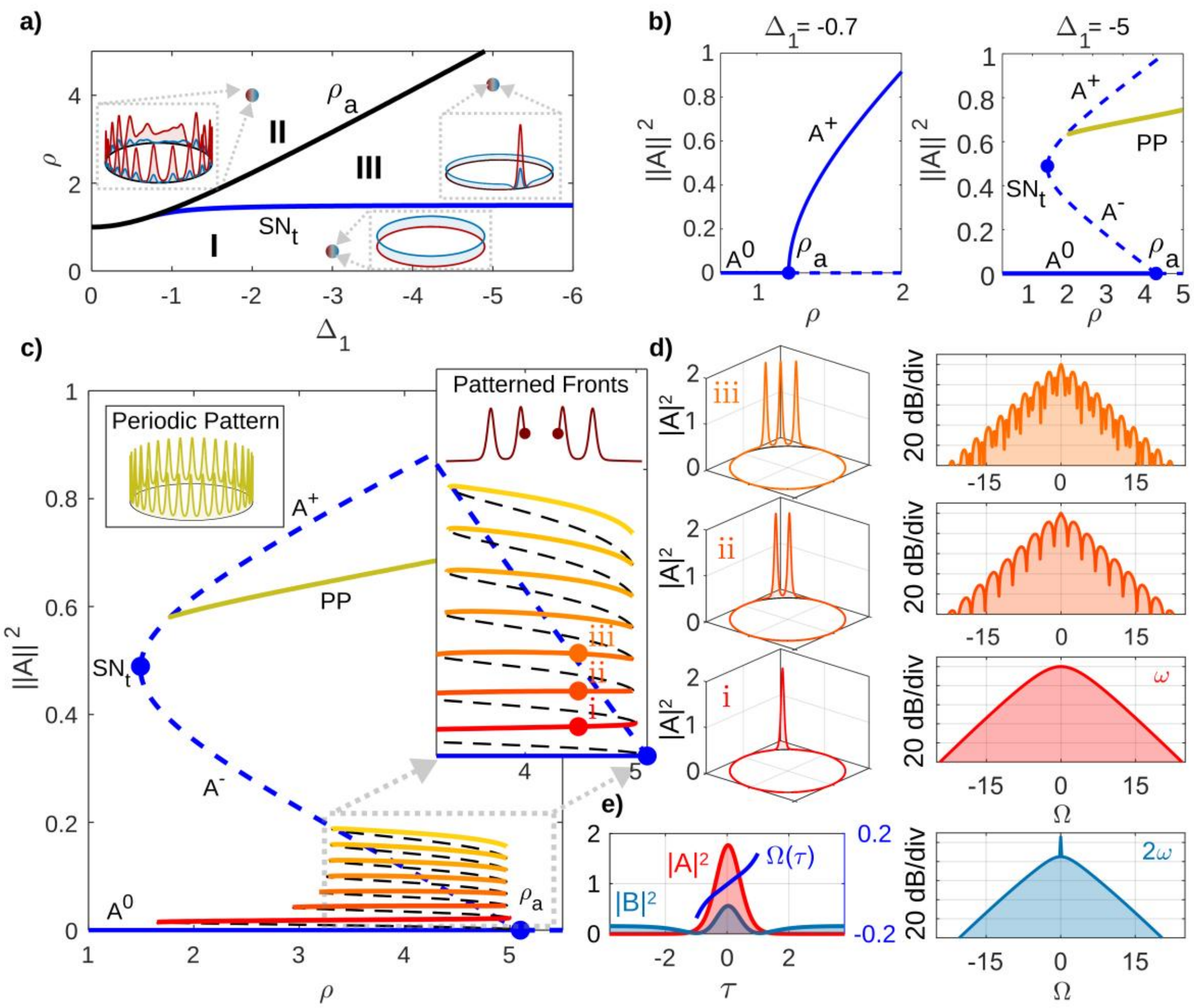

**Fig. 2 | DQS bifurcation and stability.** a) Phase diagram of homogenous intracavity solutions in the $(\Delta_1, \rho)$ parameter space. The pitchfork bifurcation $\rho_a$ (black) and turning point $SN_t$ (blue) separate the phase space into three regimes: insets show simulated pump (cyan-blue) and signal (red) states in each regime. b) Homogenous solutions at a fixed detuning of $\Delta_1 = -0.7$ (left) and $\Delta_1 = -5$ (right), with linearly stable and unstable branches shown solid and dashed, respectively. The periodic pattern (PP, dark yellow) spans the Turing bistability regime supporting DQSs. c) DQS bifurcation diagram, showing subcritical bifurcation from $A^0$ at $\rho_a$ (inset) and branches connected across Turing bistability by homoclinic snaking. d) Temporal (left) and spectral (right) profiles of states i–iii. e) Zoom-in view of state i showing signal intensity (red), instantaneous signal frequency ($\Omega(\tau)$, blue) and temporally trapped pump intensity (cyan-blue), with the pump spectrum at right.

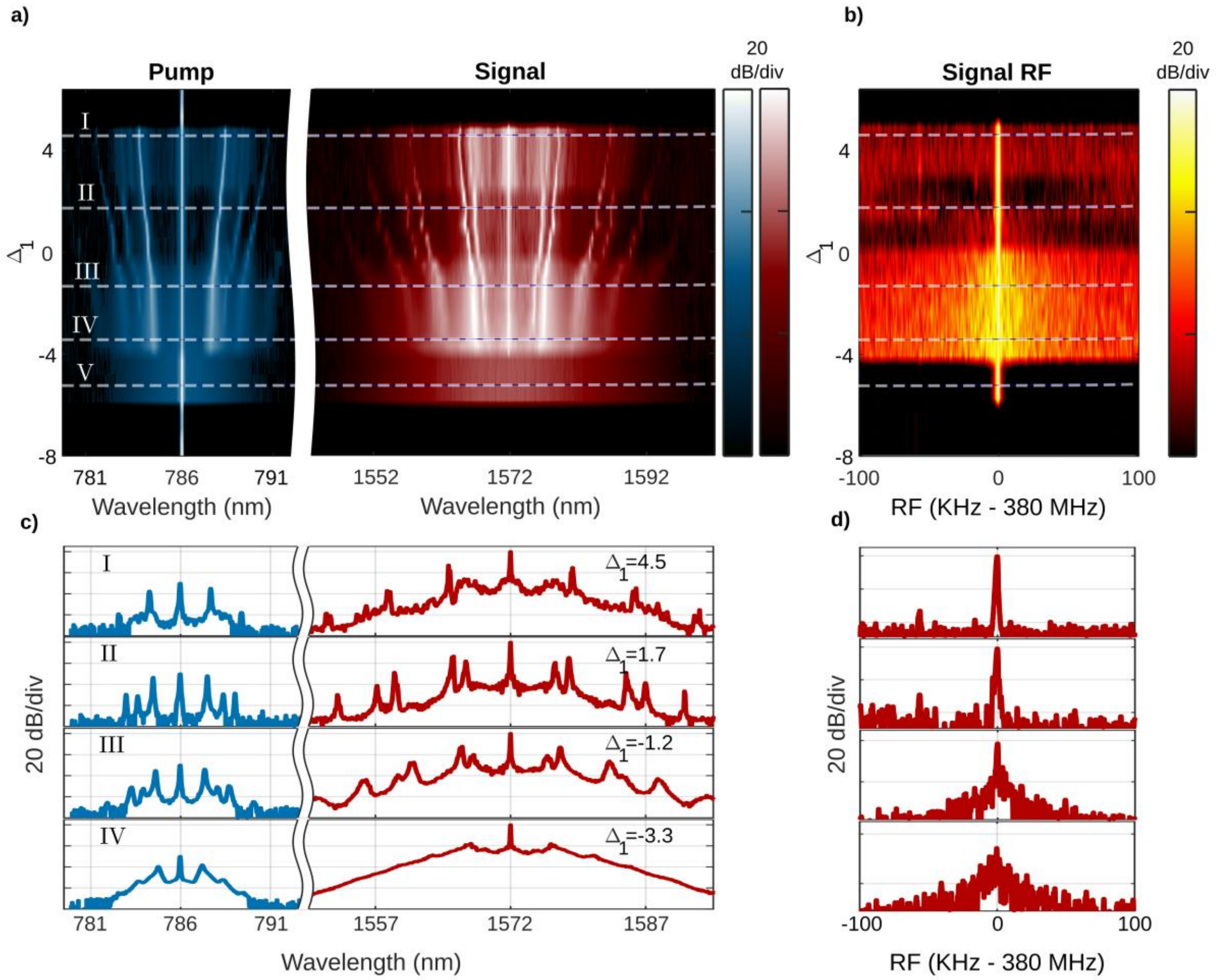

**Fig. 3 | Experimental demonstration of DQS mode-locked OPO.** a) Simultaneously measured pump (cyan-blue) and signal (red) spectra as the 600-mW continuous-wave (CW) pump is swept through cavity resonance. b) Concurrent signal radio-frequency (RF) spectra with a resolution bandwidth (RBW) of 1 kHz. c) Representative pump and signal spectra: primary (I) and secondary (II) combs, modulation instability (III), and chaotic intracavity dynamics (IV). d) Corresponding signal RF spectra indicating coherence; the broad pedestal in state IV marks the onset of intracavity chaos.

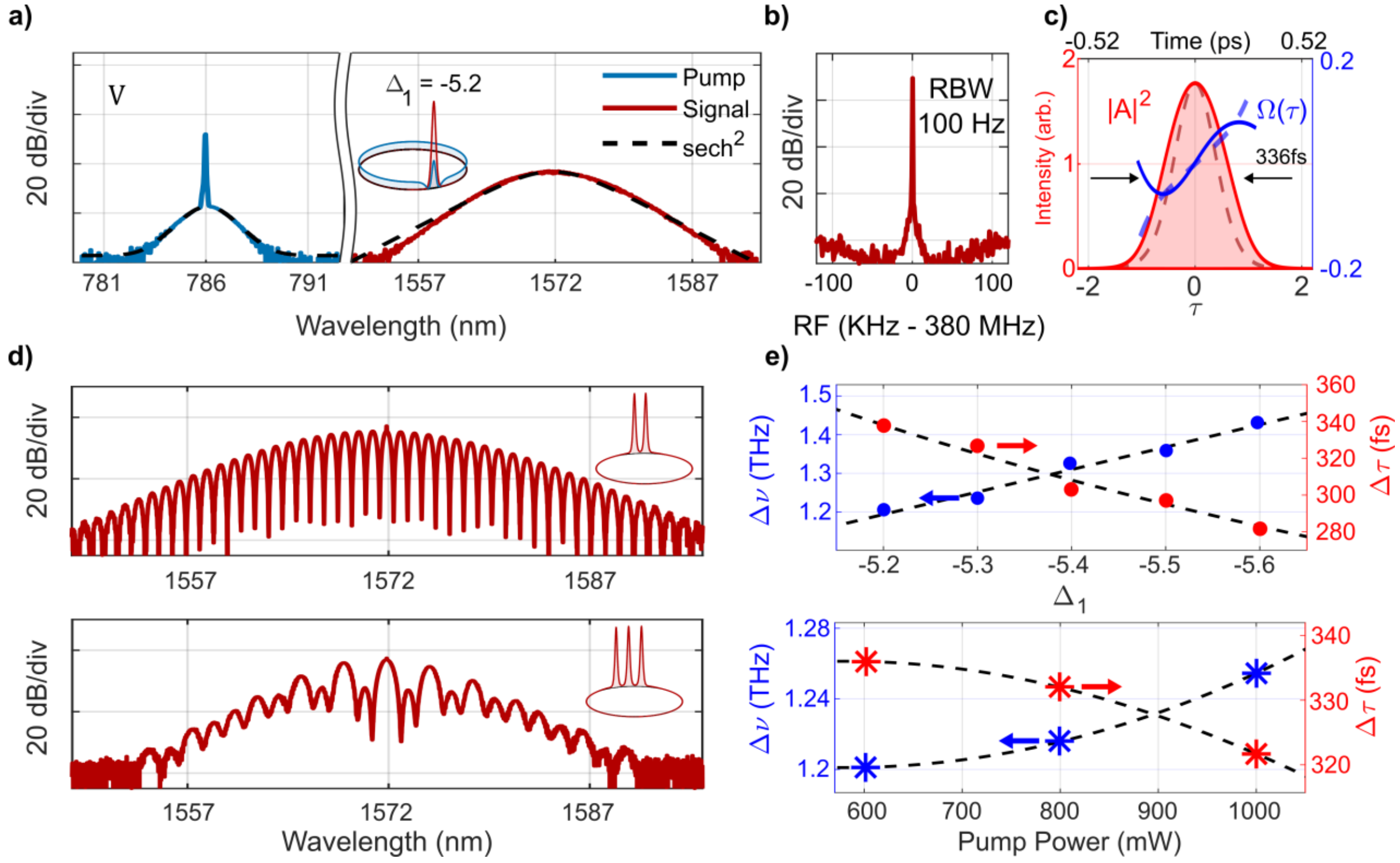

**Fig. 4 | DQS characterization.** a) Measured pump (cyan-blue) and signal (red) spectra of single DQS state V at $\Delta_1 = -5.2$ (Fig. 3), compared with predictions from stability analysis (black dashed, Fig. 2); inset shows temporal profiles. Outcoupled peak powers are 150 W (signal) and 80 W (pump), with 5% pump-to-DQS conversion efficiency. b) Signal radio-frequency (RF) spectrum with 100-Hz resolution bandwidth (RBW) and signal-to-noise ratio (SNR) exceeding 60 dB. c) Reconstructed signal pulse (red) and instantaneous frequency (blue), overlaid with the theoretical sech² profile (red dashed) from stability analysis. The measured pulse duration is 336 fs, bandwidth is 1.21 THz and time-bandwidth product is 0.407. d) Spectra of double (top) and triple (bottom) soliton molecules generated by repeated cavity sweeps (Fig. 3), with inset temporal profiles, consistent with stability analysis (Supplementary Information Section V and Supplementary Video 1). e) Single-soliton spectral bandwidth and pulse duration versus pump phase detuning (top) and pump power at $\Delta_1 = -5.2$ (bottom).

## Methods

Three of the four mirrors (M1-M3) in the bow-tie cavity are coated with high reflectivity >99.95% from 1530 nm to 1580 nm and 765 nm to 790 nm. The output coupler has ~1% transmission from 1530 nm to 1590 nm and 765 nm to 790 nm. The GVM and GVD induced by the mirror coatings are negligible. Both end facets of the 10-mm long 5% MgO-doped PPLN crystal have anti-reflection coating with high transmission at 1572 nm and 786 nm (reflectivity of 0.13% at 1572 nm and 0.09% at 786 nm). The cavity linewidth and quality factor are measured through a frequency-calibrated Mach-Zehnder interferometer (MZI). Supplementary Section XII provides the measured linewidth of 0.7 MHz, correspondingly to a loaded quality factor of ~273,000,000. The poling period of the PPLN crystal is 20.6 μm, which is designed for type-I phase-matching ($o$+$o$→$e$) at ~1572 nm. The PPLN crystal is temperature controlled with a resolution of 10 mK and the whole cavity is enclosed in a plastic box to minimize the ambient disturbance.

Supplementary Section XII provides a schematic of the experimental setup. Two external cavity diode lasers (ECDL) are phase-locked at approximately one cavity FSR (ECDL used TOPTICA CTL-1550). The CW pump provided by ECDL-1 is amplified by a 5W EDFA (IPG EAR-LP-SF) and then frequency doubled to 786nm. The SHG of the pump is coupled to a tapered amplifier (BoosTA-pro 780). The DQS pump power threshold was found to be as low as 350 mW.

To precisely control the pump phase detuning, the second ECDL is deployed as an auxiliary laser and is locked to the cavity via the Pound–Drever–Hall (PDH) method. The polarization of the PDH auxiliary laser is matched to that of the CW pump ($p$-polarization), and its power is limited to 10 mW to avoid inducing appreciable nonlinear effects. Furthermore, the CW pump and the frequency-shifted auxiliary laser counter-propagate to minimize crosstalk and are separated spectrally. The signal is output coupled via a polarization beam splitter as the generated signal is orthogonally polarized to the PDH signal.

Future implementations could be substantially simplified using offset-sideband PDH locking, which provides precise pump phase detuning control without an auxiliary laser and has been successfully applied to DKS generation [64,65]. Furthermore, nonlinear self-injection locking could enable even greater simplification by eliminating the need for active detuning control [66].

## Data availability

The data supporting the findings of this study are available in Zenodo at https://zenodo.org/records/23405139.

## Code availability

The code supporting the findings of this study are available in Zenodo at https://zenodo.org/records/23405139.

## Methods-only references

## Table of contents

# I. Theory of DQS mode-locking of CW-pumped doubly resonant optical parametric oscillation

In this section a normalized single mean field equation is derived from the Takeda mapping of the pump and signal field. This analysis derives the nonlocal nonlinear response responsible for the DQS mode-locking, a purely resonant effect caused by the interacting pump and signal fields over many round trips. We discuss this nonlocal nonlinear response in detail and describe two zones which permit soliton generation depending on the dispersion regime of the pump and signal fields. Following the analysis of this section, Section III will derive Eq. 1 of the main text used for the stability analysis. In physical units, the evolution of the pump and signal fields can be modeled by a set of coupled equations [1–3]:

$$\frac{\partial E_{s,m}}{\partial \tilde{z}} = \left[-\frac{\alpha_{c1}}{2} - i\frac{k_1''}{2}\frac{\partial^2}{\partial \tilde{\tau}^2}\right]E_{s,m} + i\kappa E_{p,m}E_{s,m}^{*}e^{-i\Delta k\tilde{z}} \qquad (S1)$$

$$\frac{\partial E_{p,m}}{\partial \tilde{z}} = \left[-\frac{\alpha_{c2}}{2} - \Delta k'\frac{\partial}{\partial \tilde{\tau}} - i\frac{k_2''}{2}\frac{\partial^2}{\partial \tilde{\tau}^2}\right]E_{p,m} + i\kappa E_{s,m}^{2}e^{i\Delta k\tilde{z}} \qquad (S2)$$

with boundary conditions:

$$E_{s,(m+1)}(0,\tilde{\tau}) = \sqrt{1-\theta_1}E_{s,m}(L,\tilde{\tau})e^{-i\delta_1} \qquad (S3)$$

$$E_{p,(m+1)}(0,\tilde{\tau}) = \sqrt{1-\theta_2}E_{p,m}(L,\tilde{\tau})e^{-i\delta_2} + \sqrt{\theta_2}E_{in}. \qquad (S4)$$

Here $E_s$ is the signal electric-field envelope and $E_p$ is the pump field envelope in units of $\sqrt{W}$ and $m$ labels the round-trip index. $E_{in}$ is the CW driving field, $\alpha_{c1,2}$ is the loss per unit length, $\Delta k$ is the wave-vector mismatch, $\Delta k'$ is the group velocity mismatch (GVM) between the pump and signal fields, and $k_{1,2}''$ are the group-velocity dispersions (GVDs), $L$ is the cavity length, $\theta_{1,2}$ are the coupler transmission coefficients and $\delta_{1,2}$ are the phase detunings of the signal and pump respectively. Higher-order dispersion and nonlinearity are both neglected for simplicity. $\kappa = \sqrt{2}\omega_0 d_{eff}/\sqrt{A_{eff}c^3 n_s n_p^2 \varepsilon_0}$ is the second-order nonlinear coupling coefficient, where $2\omega_0$ is the pump carrier frequency, $d_{eff}$ is the effective second-order nonlinear coefficient, $A_{eff}$ is the effective mode area, $c$ is the speed of light, $\varepsilon_0$ is the vacuum permittivity, and $n_{p,s}$ are the linear refractive indices.

In what follows, we assume the combined effects of nonlinearity and dispersion over a single round trip are "weak" such that the fields $E_s$ and $E_p$ do not significantly change. Under these conditions, the coupled Equations (S1)-(S4) can be reduced to two coupled mean-field equations following similar techniques from references [4,5] where we integrate over the cavity length, $L$

$$E_{s,m}(L,\tilde{\tau}) = E_{s,m}(0,\tilde{\tau}) - L\left[-\frac{\alpha_{c1}}{2} - i\frac{k_1''}{2}\frac{\partial^2}{\partial \tilde{\tau}^2}\right]E_{s,m}(0,\tilde{\tau}) + i\kappa E_{p,m}(0,\tilde{\tau})E_{s,m}^{*}(0,\tilde{\tau})e^{-i\xi}\mathrm{sinc}(\xi) \qquad (S5)$$

$$E_{p,m}(L,\tilde{\tau}) = E_{p,m}(0,\tilde{\tau}) - L\left[-\frac{\alpha_{c2}}{2} - i\frac{k_2''}{2}\frac{\partial^2}{\partial \tilde{\tau}^2}\right]E_{p,m}(0,\tilde{\tau}) + i\kappa {E_{s,m}}^{2}(0,\tilde{\tau})e^{i\xi}\mathrm{sinc}(\xi). \qquad (S6)$$

Here we have defined the normalized phase-mismatch as $\xi = \Delta kL/2$. We substitute Eq. (S5) into Eq. ($S3$) and Eq. ($S6$) into Eq. ($S4$) and take the field phase-detunings and dispersive terms to first order. By neglecting these higher order terms, we find

$$E_{s,(m+1)}(L,\tilde{\tau}) - E_{s,m}(0,\tilde{\tau}) = \left(-\alpha_1 - i\delta_1 - i\frac{k_1''L}{2}\frac{\partial^2}{\partial \tilde{\tau}^2}\right)E_{s,m}(0,\tilde{\tau}) + i\kappa E_{p,m}(0,\tilde{\tau})E_{s,m}^{*}(0,\tilde{\tau})e^{-i\xi}\mathrm{sinc}(\xi) \qquad (S7)$$

$$E_{p,(m+1)}(L,\tilde{\tau}) - E_{p,m}(0,\tilde{\tau}) = \left(-\alpha_2 - i\delta_2 - \Delta k'L\frac{\partial}{\partial \tilde{\tau}} - i\frac{k_2''L}{2}\frac{\partial^2}{\partial \tilde{\tau}^2}\right)E_{p,m}(0,\tilde{\tau}) + i\kappa {E_{s,m}}^{2}(0,\tilde{\tau})e^{i\xi}\mathrm{sinc}(\xi) + \sqrt{\theta_2}E_{in}. \qquad (S8)$$

Here $\alpha_{1,2} = (\theta_{1,2} + \alpha_{c1,2}L)/2$ is the total round-trip loss. We next introduce a slow-time variable, $\tilde{t}$, which describes the evolution of the fields over each pass at the boundary. We then define the intracavity envelopes for the signal and pump as $\tilde{\mathrm{E}}_s(\tilde{t},\tilde{\tau})$ and $\tilde{\mathrm{E}}_p(\tilde{t},\tilde{\tau})$ and replace the roundtrip index, $m$, with the slow-time variable

$$\tilde{\mathrm{E}}_s(\tilde{t} = mt_r,\tilde{\tau}) = E_{s,m}(z=0,\tilde{\tau}), \qquad m = 0,1,2,\ldots, \qquad (S9)$$

$$\widetilde{\mathrm{E}}_{\mathrm{p}}(\tilde{t} = mt_r, \tilde{\tau}) = E_{p,m}(z = 0, \tilde{\tau}), \qquad m = 0,1,2, \dots. \qquad (S10)$$

Here the cavity roundtrip time is defined $t_r$. We next approximate the finite difference $\left[E_{s,(m+1)} - E_{s,m}\right]/t_r$ with a partial derivative $\partial\widetilde{\mathrm{E}}_s \,/\, \partial\tilde{t}$ and with a similar approximation for the pump field we obtain a set of two coupled mean field equations:

$$t_r \frac{\partial \widetilde{\mathrm{E}}_s}{\partial \tilde{t}} = \left(-\alpha_1 - i\delta_1 - i\frac{k_1''L}{2}\frac{\partial^2}{\partial\tilde{\tau}^2}\right)\widetilde{\mathrm{E}}_s + i\kappa\widetilde{\mathrm{E}}_p\widetilde{\mathrm{E}}_s^* e^{-i\xi}\operatorname{sinc}(\xi) \qquad (S11)$$

$$t_r \frac{\partial \widetilde{\mathrm{E}}_p}{\partial \tilde{t}} = \left(-\alpha_2 - i\delta_2 - \Delta k' L\frac{\partial}{\partial\tilde{\tau}} - i\frac{k_2''L}{2}\frac{\partial^2}{\partial\tilde{\tau}^2}\right)\widetilde{\mathrm{E}}_p + i\kappa{\widetilde{\mathrm{E}}_s}^2 e^{i\xi}\operatorname{sinc}(\xi) + \sqrt{\theta_2}E_{in}. \qquad (S12)$$

Eq. $(S11)$ and Eq. $(S12)$ can be simplified to dimensionless forms following the substitutions $t = \alpha_1\tilde{t}/t_r, \tau = \tilde{\tau}(2\alpha_1/|k_1''|L)^{1/2}$, $\mathrm{a} = \widetilde{\mathrm{E}}_s\kappa L/\alpha_1$, $\mathrm{b} = \widetilde{\mathrm{E}}_p\kappa L/\alpha_1$, $\Delta_{1,2} = \delta_{1,2}/\alpha_1$, $d = \Delta k'(2L/\alpha_1|k_1''|)^{1/2}, \beta_1 = sign(k_1''), \beta_2 = k_2''/|k_1''|, \alpha = \alpha_2/\alpha_1, d = \Delta k'\sqrt{2L/\alpha_1|k_1''|}$ and $S = \sqrt{\theta_2}E_{in}\kappa L/\alpha_1^2$ to get a normalized set of coupled mean field equations

$$\frac{\partial \mathrm{a}}{\partial t} = \left(-1 - i\Delta_1 - i\beta_1\frac{\partial^2}{\partial\tau^2}\right)\mathrm{a} + i\mathrm{b}\mathrm{a}^* e^{-i\xi}\operatorname{sinc}(\xi) \qquad (S13)$$

$$\frac{\partial \mathrm{b}}{\partial t} = \left(-\alpha - i\Delta_2 - id\frac{\partial}{\partial\tau} - i\beta_2\frac{\partial^2}{\partial\tau^2}\right)\mathrm{b} + i\mathrm{a}^2 e^{i\xi}\operatorname{sinc}(\xi) + \mathrm{S}. \qquad (S14)$$

In the main text the normalized frequency is also used to quantify the soliton spectral bandwidth. This is defined in normalized units where $\Omega = \nu/(2\alpha_1/|k_1''|L)^{1/2}$ and $\nu$ is in unit of Hz and corresponds to the frequency offset from the carrier.

For a wide range of parameters, the pump field will evolve slowly such that $\partial\mathrm{b}/\partial t = 0$ and the pump field can be solved for explicitly in the Fourier domain,

$$\mathcal{F}\{\mathrm{b}\} = \frac{i\mathcal{F}\{\mathrm{a}^2\}e^{i\xi}\operatorname{sinc}(\xi) + \mathrm{S}\delta_0(\Omega)}{-\alpha - i\Delta_2 - id\Omega - \beta_2\Omega^2}, \qquad (S15)$$

where $\mathcal{F}$ denotes the Fourier transform and $\delta_0(\Omega)$ is the Kronecker delta function.

Substituting Eq. $(S15)$ into Eq. $(S13)$ we find the evolution of the signal field obeys the single mean field equation

$$\frac{\partial \mathrm{a}}{\partial t} = \left(-1 - i\Delta_1 - i\beta_1\frac{\partial^2}{\partial\tau^2}\right)\mathrm{a} - \mathrm{a}^*\left(\mathrm{a}^2 \otimes J(\tau)\right) + S\mathrm{a}^* e^{2i\psi}. \qquad (S16)$$

Here $\psi = \xi/2 + \tan^{-1}(-\Delta_2/\alpha)/2$ and the pump field is dynamically slaved to the signal field as

$$\mathrm{b} = (-\mathrm{a}^2 \otimes J(\tau) + S)e^{i\tan^{-1}(\Delta_2/\alpha)} \qquad (S17)$$

$\otimes$ denotes the convolution operator and the non-local response is then found to be:

$$\mathcal{F}\{J(\tau)\} = \hat{J}(\Omega) = \frac{\operatorname{sinc}^2(\xi)}{\alpha + i(\Delta_2 - d\Omega - \beta_2\Omega^2)}. \qquad (S18)$$

Section XII provides a table summarizing the definitions and values of all parameters corresponding to the experimental conditions.

To elucidate the different effects of this dispersive nonlinear response, we separate its frequency domain response into the real, $X(\Omega)$, and imaginary, $Y(\Omega)$. In the perfect phase-matched condition, we set $\xi = 0$ and $\alpha = 1$ which allows us to express the nonlinear response as

$$X(\Omega) = [1 + (\Delta_2 - d\Omega - \beta_2\Omega^2)^2]^{-1} \qquad (S19)$$

$$Y(\Omega) = [-(\Delta_2 - d\Omega - \beta_2\Omega^2)]X(\Omega). \tag{S20}$$

Under these perfect phase-matching conditions the pump and signal phase-detuning are related, $\Delta_2 = 2\Delta_1$ (see Section II). Together, $X(\Omega)$ and $Y(\Omega)$ are representative of a dispersive effective two photon effect (ETPA) and a dispersive effective Kerr nonlinearity (EKN). Importantly both the ETPA and EKN are solely driven by the design of the pump parameters. Fig. S1a illustrates the EKN and ETPA as a function of the signal detuning and Fig. S1b illustrates the nonlocal-response on an ultrafast optical pulse, where we convolved the inverse Fourier transformed nonlinearity with a 250 fs transform limited gaussian. Fig. S1a is plotted in normalized units corresponding to the table of simulation and cavity parameters provided in Section XII. The temporal bandwidths of the test pulse are illustrated as the dashed lines.

Two distinct regimes can be identified from Fig. S1a, the upper zone, where $\beta_2\Delta_1 > 0$, is characterized by two symmetric resonant spectral anomalies while the lower zone, where $\beta_2\Delta_1 < 0$, is characterized by a smooth frequency response with increasing bandwidth at larger $|\Delta_1|$. The solutions of the lower nonlinear response zone will exhibit a smooth profile compared to the upper zone where the narrowband resonant peaks result in disruptive spectral modulation. In our cavity, where $\beta_2 > 0$ and primarily governed by the normal material dispersion of the PPLN, clean lower-zone DQS with bichromatic sech$^2$ spectra is expected when $\Delta_1 < 0$.

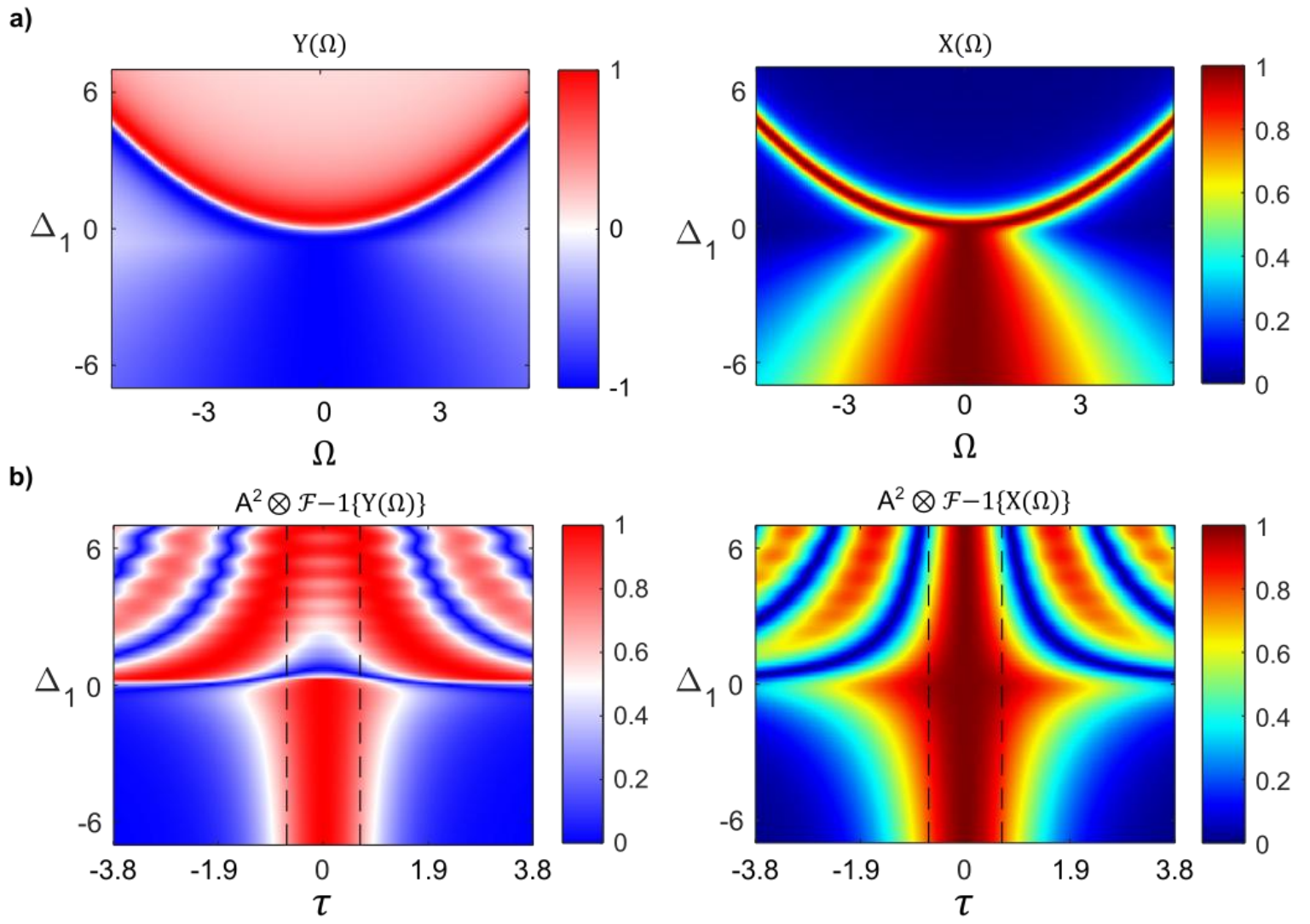

**Fig. S1: Nonlocal Nonlinear Response:** The nonlinear response can be separated into a) the imaginary (left) and real (right) components which mimic a dispersive effective Kerr nonlinearity (EKN) and an effective two photon absorption (ETPA). These effects are separated into to the upper zone, $\Delta_1\beta_2 > 0$, where phase anomalies in the ETPA cause disruptive pulse distortions compared to the lower zone, $\Delta_1\beta_2 < 0$, where the response is a smooth band-pass filter. The effects of these two nonlinearities are illustrated in b) where the ETPA and EKN response are shown in the time domain by convolving with a 250 fs test pulse which is plotted in normalized units, the dash lines illustrate the test pulse duration. The simulation parameters are provided in Section XII. The magnitudes of both a) and b) are normalized such that the peak at each detuning value is one.

The upper zone's two resonance-enhanced ETPA peaks are associated with nonlinear phase-anomalies which resemble the narrowband perturbations of mode-crossings in conventional DKS formation. These narrowband resonant peaks are responsible for the comb lines that are initially generated in the red detuned region of our experimental sweep of the OPO cavity (see Fig. 3a of the main text). These anomalies are symmetric about zero at $\Omega = \pm\sqrt{2\Delta_1/\beta_2}$. The convolution with the test pulse reveals long oscillatory tails in the pulse shape, caused by these narrow band perturbations. This long-range interaction and pulse perturbation will, in general, lead to pulse destabilization. A simple strategy to avoid these detrimental effects is to keep the resonant peaks well separated compared to the pulse bandwidth. Ideally, $2\sqrt{2\Delta_1/\beta_2} \gg \pi/\Delta T$ where $\Delta T$ is the normalized pulse width. This condition can be achieved by engineering a small $\beta_2$ or operating at large $\Delta_1$. Indeed, it has been theoretically shown that upper zone solitons can be sustained given that the cavity dispersion is such that $\beta_1\beta_2 < 0$ and $\Delta_1\beta_2 > 0$ where the EKN is engineered to be opposite of the dispersion [3,6]. In these cases, the dispersive nonlinear response results in a different spectral envelope compared to the lower branch sech$^2$ spectra which reflects the narrowband phase anomalies. Table S1 summarizes the different conditions in which dissipative quadratic solitons may exist.

| Soliton Regime | $\beta_1$ | $\beta_2$ | $\Delta_1$ |
|---|---|---|---|
| Upper zone | - | + | + |
| Upper zone | + | - | - |
| Lower zone | - | - | + |
| Lower zone | + | + | - |

**Table S1***:* Bright DQS existence in doubly resonant degenerate optical parametric oscillator.

In the case of our all-normal dispersion cavity, $\beta_1\beta_2 > 0$, the dispersion of the nonlinear response is a smooth bandpass and the formation of bright solitons, as was shown in the main text, are allowed in the lower branch, $\Delta_1\beta_2 < 0$. In this lower zone of the nonlinear response, no narrowband resonance phenomenon is observed, and the frequency response is a smooth band pass filter where the bandwidth approaches $2\sqrt{2(\sqrt{2}-1)|\Delta_1/\beta_2|}$. This much smoother and localized nonlinear response does not have long-range interactions which allows for cleaner pulse shape and the sech$^2$ spectra which produces the results of the main text (see Fig. 2 and Fig. 4 of the main text). Of note, either lower or upper branch solitons can be supported given that the engineered nonlinearity – controlled by the pump phase-detuning – is set to balance the cavity dispersion.

In the limit where the dispersion of the nonlinearity can be ignored, the response can be simplified to constant values in frequency

$$\kappa^2 L\hat{J}(0) \equiv \alpha_{ETPA} + i\gamma_{eff}. \qquad (S21)$$

Here $\gamma_{eff} = \kappa^2 LY(0)$ and $\alpha_{ETPA} = \kappa^2 LX(0)$ and corresponds to the EKN and ETPA parameters as outlined in the operating principle of the main text. From Eq. S7 and S8, $Y(0)/X(0) = \gamma_{eff}/\alpha_{TPA} = \Delta_2$ meaning that the EKN to ETPA ratio is directly proportional to the pump resonant detuning and the EKN will dominate over the ETPA at $\Delta_2 \gg 1$. Since $\alpha_{ETPA}$ dictates the amount of reconversion of the signal back into the pump field, at larger detuning's, the relative energy of the signal to the pump will increase quadratically with the detuning. Additionally, $\gamma_{eff}$ is inversely proportional to $\Delta_2$ such that the peak power of the soliton will increase quadratically with larger detuning.

Besides doubly resonant configurations, singly resonant cavities provide an additional pathway towards an engineered and tunable nonlinearity. Table S2 provides a comparison of the EKN and ETPA coefficients depending on the resonance condition. Of note, the doubly resonant condition provides additional dependence on the pump detuning compared to the singly resonant configuration and will minimize the ETPA coefficient at large pump detuning's. Comparatively the singly resonant case is entirely controlled by the phase-mismatch such that the EKN to ETPA ratio is maximized at $\xi = \pi$.

| Cavity configuration | Effective Kerr Nonlinearity (EKN) $\gamma_{eff}$ | Effective Two-Photon Absorption (ETPA) $\alpha_{ETPA}$ |
|---|---|---|
| **Singly resonant (SR)** | $\gamma_{eff} = \kappa^2 L\frac{1 - sinc(2\xi)}{2\xi}$ | $\alpha_{ETPA} = \kappa^2 L\frac{sinc^2(\xi)}{2}$ |
| **Doubly resonant (DR)** | $\gamma_{eff} = \kappa^2 Lsinc^2(\xi)\frac{\Delta_2\alpha_1}{(\alpha\alpha_1)^2 + (\Delta_2\alpha_1)^2}$ | $\alpha_{ETPA} = \kappa^2 Lsinc^2(\xi)\frac{\alpha\alpha_1}{(\alpha\alpha_1)^2 + (\Delta_2\alpha_1)^2}$ |

**Table S2:** The EKN and ETPA coefficients resulting from the different cavity resonant conditions.

The main differences between DQS-modelocked cavity enhanced SHG cavities and DQS-modelocked optical parametric oscillators arise from the parametric process which the pump directly drives. In OPO, the pump drive term directly seeds an OPG process allowing the pump frequency to satisfy an essentially continuous spectrum of energy conserving down conversions. Meanwhile, in cavity enhanced SHG processes, the pumping may only directly drive the frequency doubled - energy conserved - up conversion. Physically, this is where the parametric drive term originates in Eq. (S16). In accordance with this analysis, the parametric driving term is crucial to maximize efficient energy transfer from the pump to signal field.

## II. Phase-matching and resonant detuning

In this section we describe how the doubly resonant and phase-matched condition is achieved in our quadratic nonlinear resonator. To successfully run the doubly resonant optical parametric oscillator, with low threshold, both the signal detuning $\delta_1$ and pump detuning $\delta_2$ should be limited to a small value such that they are resonant close to the cavity resonance peaks. Therefore, the wave numbers ($k_{1,2}$) should satisfy,

$$k_1 L - \delta_1 = m \cdot 2\pi \ (m \in \boldsymbol{Z})$$
$$k_2 L - \delta_2 = m \cdot 2\pi \ (m \in \boldsymbol{Z})$$
$$\Delta k = k_2 - 2k_1 - 2\pi/\Lambda$$

$$\delta_2 = 2\delta_1 + \Delta k L + 2\pi L/\Lambda + p \cdot 2\pi \ (p \in \boldsymbol{Z}) \qquad (S22)$$

where $\Lambda$ is the artificial quasi-lattice period. In a homogenous cavity and under the phase matched condition, $\Delta k L = 0$ ($\Lambda = \infty$), and $\delta_2 = 2\delta_1$ [4]. For quasi-phase-matching case, when $\Delta k L + 2\pi L/\Lambda = p \cdot 2\pi \ (p \in \boldsymbol{Z})$, will also yield $\delta_2 = 2\delta_1$. In this case, we can conveniently adjust the phase matching condition $\Delta k L$ through the quasi-lattice period $\Lambda$ and always satisfy the condition that $\delta_2 = 2\delta_1$. In the situation of the main text, we first achieve $\Delta k L = 0$ and then the cavity length is adjusted to be an integer multiple of the quasilattice period which is achieved by adjusting the translation stage of the cavity.

## III. Stability analysis of the doubly resonant optical parametric oscillator

In this section we conduct a stability analysis of the DR-DOPO system. Our analysis reveals the existence of the DQS mode-locking regime which was described in the main text. We expand this analysis to reveal the stable and unstable existence regime of the cavity solutions.

The normalized single mean field equation, Eq. $(S16)$, does not lend itself well to conducting this stability analysis due to the detuning dependent nonlinear response and the complex exponential in the parametric drive term. To proceed, we first rescale the normalized field such that $\mathrm{a} = \mathsf{A} e^{i\psi}\sqrt{\alpha(1 + (\Delta_2/\alpha)^2)}$, and similarly for the pump field, which will allow Eq. $(S16)$ and Eq. $(S17)$ to be rewritten:

$$\partial_\tau \mathsf{A} = -(1 + i\Delta_1)\mathsf{A} - i\beta_1 \partial_\tau^2 \mathsf{A} - \mathsf{A}^*(\mathsf{A}^2 \otimes \bar{J}(\tau)) + \rho \mathsf{A}^* \qquad (S23)$$

and

$$\mathsf{B} = \left(-\mathsf{A}^2 \otimes \bar{J}(\tau) + \rho\right) e^{i\tan^{-1}(\Delta_2/\alpha)} \Big/ \sqrt{\alpha(1 + \Delta_2^2/\alpha^2)} \qquad (S24)$$

where the rescaled parametric drive term to $\rho = S/\alpha\sqrt{1 + (\Delta_2/\alpha)^2}$. This rescaling requires us to rewrite the nonlinearity to:

$$\bar{J}(\tau) = \mathrm{J}(\tau)[\alpha(1 + (\Delta_2/\alpha)^2)] = \ \mathrm{sinc}^2(\xi)\, \mathcal{F}\left\{ \frac{\alpha(1 + (\Delta_2/\alpha)^2)}{\alpha + i(\Delta_2 - d\Omega - \beta_2\Omega^2)} \right\}. \qquad (S25)$$

Eq. $(S23)$-$(S25)$ are identical to Eq. (1)-Eq. (3) of the main text. Section XII provides a table of all parameters and their values used in the following calculations. The relationship between all of the physical, normalized, and rescaled values are also provided. With this rescaled mean field and non-local nonlinearity established, we now proceed theoretically with the generation of the DQS using a rigorous stability analysis of the system's steady state solutions.

The steady state solutions can be a variety of different solutions including homogenous states, periodic patterns, local structures, or domain walls. In this work we look for and define localized high peak power solutions as bright dissipative quadratic solutions. All stationary solutions must satisfy the relation $\partial_\tau \mathsf{A} = 0$

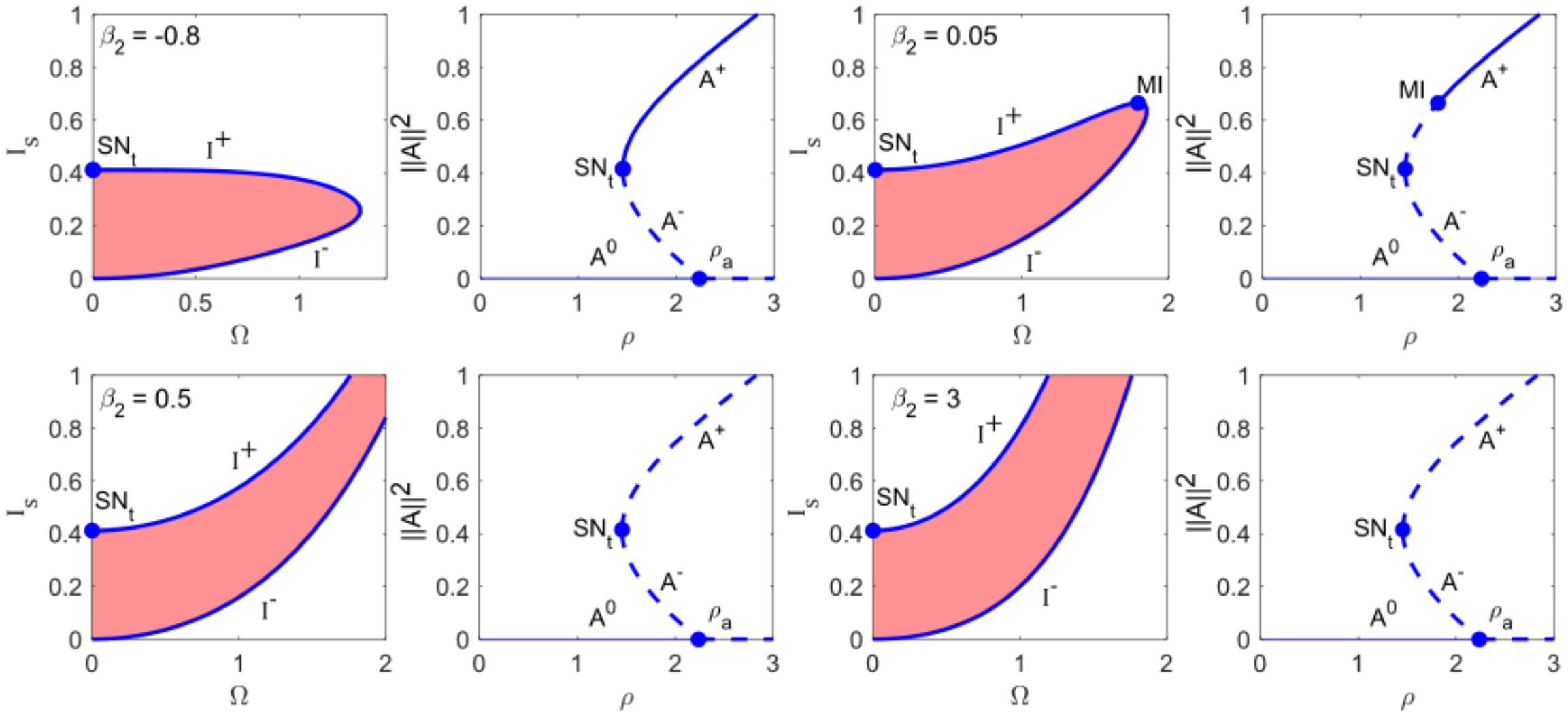

**Fig. S2 Stability analysis of the Homogenous cavity solutions with varying pump to signal dispersion ratio:** The marginal instability curve defines the edges of the modulational instability (MI) band for the non-trivial homogenous solutions, $\mathrm{A}^{\pm}$. All figures are plotted for identical detuning and in the normal dispersion regime, $(\Delta_1, \beta_1) = (-2,1)$ and the pump to signal dispersion ratio, $\beta_2$, is varied. The unstable region is defined by the red-shaded area. The MI threshold occurs at the maximum of the curve at the critical frequency of which the maximum occurs indicated by the point MI. Therefore, the homogenous solutions which possess less energy than this threshold will be unstable. This is best illustrated by comparing $\beta_2 = -0.8$ where $\mathrm{A}^{+}$is always stable with $\mathrm{A}^{0}$ while at $\beta_2 = 0.05$ $\mathrm{A}^{+}$ is only stable beyond $\rho > \sim 1.8$. For $\beta_2 = 0.5$ and $\beta_2 = 3$ $\mathrm{A}^{+}$ is never stable in the same region as $\mathrm{A}^{0}$.

### a. Homogenous solutions

The homogenous states of Eq. $(S23)$ have been studied rigorously in [7,8] thus we turn the reader's attention to these references for a thorough derivation. Here we present the main results and establish them in the context of our resonator. The homogeneous solutions of Eq. $(S23)$ must satisfy the equation

$$0 = -(1 + i\Delta_1)\mathrm{A} - (1 + i\Delta_2)\ |\mathrm{A}|^2\mathrm{A} + \rho\mathrm{A} \qquad (S26)$$

This yields three solutions, the trivial solution $\mathrm{A}^0 = 0$ and two non-trivial solutions: $\mathrm{A}_s^{\pm} = |\mathrm{A}^{\pm}|e^{i\phi}$ with amplitude

$$|\mathrm{A}^{\pm}|^2 = \frac{\left((\Delta_1\,\Delta_2/\alpha - 1) \pm \sqrt{(1 + (\Delta_2/\alpha)^2)\rho^2 - (\Delta_2/\alpha + \Delta_1)^2}\right)}{\mathrm{sinc}^2(\xi)(1 + (\Delta_2/\alpha)^2)} \qquad (S27)$$

and phase

$$\phi^{\pm} = \frac{\mathrm{acos}\left(\left[|\mathrm{A}^{\pm}|^2 + 1\right]/\rho\right)}{2}. \qquad (S28)$$

As was shown in the main text $\mathrm{A}^{+}$ arises subcritically from the trivial solution when $\Delta_2\Delta_1/\alpha < 1$ at a pitchfork bifurcation occurring at

$$\rho_{\mathrm{a}} = \sqrt{1 + \Delta_1^2}. \qquad (S29)$$

When $\Delta_2\Delta_1/\alpha > 1$, both $\mathrm{A}^{+}$ and $\mathrm{A}^{-}$ coexist and connect at a fold or turning point located at

$$\mathrm{SN_t} = (\Delta_2/\alpha + \Delta_1)/\sqrt{1 + (\Delta_2/\alpha)^2}$$

To better understand the stability of these homogenous solutions, $\mathrm{A}_s^{\pm}$, we insert the ansatz

$$\mathrm{A}(\mathrm{t}, \Omega) = \mathrm{A}_s^{\pm} + \epsilon\xi e^{\sigma\mathrm{t} + \mathrm{i}\Omega\tau} + c.c. \qquad (S30)$$

into Eq. $(S23)$. Which describes a small modulation around the homogenous solution. The growth rate, $\sigma$, and eigenvector, $\xi$, of the perturbation is associated with the linearization of Eq. $(S23)$ at an order of $\epsilon$. The solutions are linearly stable at a given $\Omega$ if the growth rate of the perturbation satisfies $\mathrm{Re}[\sigma] < 0$ and unstable otherwise. Plugging this ansatz into Eq. $(S23)$ and keeping only the linear terms in $\epsilon$, this growth rate satisfies the quadratic relationship

$$\sigma^2 + a_1\sigma + a_0 = 0 \tag{S31}$$

Where the coefficients of this quadratic equation is defined:

$$a_1 = 2(1 + 2\mathrm{I_s}Re\{\bar{J}(\Omega)\}) \tag{S32}$$

and

$$a_0 = c_2\mathrm{I_s^2} + c_1\mathrm{I_s} + c_0. \tag{S33}$$

Where, $\mathrm{I}_s = |\mathrm{A}|^2$is the normalized and rescaled signal power. Eq. $(S33)$ is quadratic with signal power and has coefficients defined:

$$c_2 = 4(|\bar{J}(\Omega)|^2), \tag{S34}$$

$$c_1 = 4(Re\{\bar{J}(\Omega)\} - (\beta_1\Omega^2 - \Delta_1)Im\{\bar{J}(\Omega)\}), \tag{S35}$$

and

$$c_0 = \Omega^4 - 2\beta_1\Delta_1\Omega^2. \tag{S36}$$

To start we set $\mathrm{I_s} = 0$ which corresponds to the solutions for the growth rate for the trivial homogenous solution, $\mathrm{A}^0$. Under this condition, the growth rate is simply

$$\sigma = -1 \pm \sqrt{\rho^2 - (\Delta_1 - \beta_1\Omega^2)^2}. \tag{S37}$$

Therefore, the trivial solution becomes unstable when $\rho^2 > 1 + (\Delta_1 - \beta_1\Omega^2)^2$. The critical frequency, $\Omega_c$, at which this instability will produce the maximal modulation instability (MI) gain, will correspond to where the square root is maximized given by the relationship

$$\Omega_c^2 = \frac{\Delta_1}{\beta_1}.$$

Therefore, the trivial solution becomes unstable for $\rho > 1$ at $\Omega_c$. This establishes two conditions for the stability of $\Omega_c$ when $\rho > 1$, depending on the dispersion regime of the signal. In the normal dispersion, $\beta_1 = 1$, $\mathrm{A}^0$ is unstable when $\Delta_1 > 0$ and in the anomalous regime, $\beta_1 = -1$, $\mathrm{A}^0$ is unstable when $\Delta_1 < 0$.

On the other hand, $\mathrm{A}^0$ will always be unstable at $\Omega = 0$ when $\rho > \rho_a$, regardless of the sign of $\Delta_1$. This of course corresponds to the OPO threshold i.e. the bifurcation of $\mathrm{A}^+$ in the $(\Delta, \rho)$-plane. Notably, the instability of $\mathrm{A}^0$ does not depend on the dispersion of the pump field.

Now that we have established the stability of the trivial solution, we proceed with the linear stability of the nontrivial homogenous states, $\mathrm{A}^\pm$ . The stability of these solutions is less trivial and do not possess exact analytical expressions. Therefore, the characterization of the stability of these states can be conducted by means of the marginal instability curve (MIC) obtained from the condition that the growth rate of the perturbation is zero, i.e. solutions which correspond to $\mathrm{Re}[\sigma] = 0$. This condition defines the boundary of stable and unstable non-trivial solutions. By setting $\sigma = 0$, the solution of the MIC follows the following equation quadratic in signal powers, $\mathrm{I_s}$ [9]:

$$c_2\mathrm{I_s^2} + c_1\mathrm{I_s} + c_0 = 0. \tag{S38}$$

This leads to two solution branches of the MIC

$$\mathrm{I_s^\pm} = \frac{-c_1 \pm \sqrt{c_1^2 - 4c_2c_0}}{2c_2}. \tag{S39}$$

These two branches define the edge of the MI band. The instability of the solutions will therefore occur at the maximum of these branches where

$$\frac{dI_s}{d\Omega} = 0.$$

The MIC for a variety of pump to signal dispersion relations are provided in Fig. S2 where unstable powers exist in the red-shaded region. The trivial and non-trivial solutions for each condition are plotted as solid (dashed) blue lines for stable (unstable) solutions. The MI point illustrated in Fig. S2 defines the threshold of the instability set by the maxima of the red-shaded region. In all conditions illustrated in Fig. S2, the detuning of the signal is fixed, and the signal dispersion regime is set to normal such that $(\Delta_1, \beta_1) = (-5,1)$.

The role of the pump to signal dispersion ratio is to adjust the threshold of the stable region to larger and larger signal powers. This is best illustrated by the increase in the MI point with changing sign of the pump dispersion where in the anomalous pump dispersion regime, $\beta_2 < 0$, $A^+$ can exist stably with $A^0$. This contrasts with the normal pump dispersion regime, $\beta_2 > 0$, where $A^+$ can experience MI gain within the stable $A^0$ region. Further increasing $\beta_2$ adjusts the MI point to occur at larger $I_s$, corresponding to a larger $\rho$. In this regime stable $A^+$ never coexist with the stable $A^0$ nor the unstable $A^-$ region. It is in this regime, $\rho < \rho_a$, where the cavity may possess stable Turing pattern solutions. As was demonstrated in the main text and as will be described in the next section, it is the bistability of the stable trivial and stable Turing pattern solutions where the cavity hosts DQS solutions.

## IV. Numerical path continuation in $||A||^2$-$\rho$ parameter space

Within the instability regime of $A^{\pm}$, where $A^0$ is stable, Turing states can emerge subcritically from the non-trivial homogenous solution. Bistability in this regime occurs between the Turing and trivial solution where patterned fronts emerge connecting stable states as described in the main text. Under these conditions, the locking of these patterned fronts lead to the generation of the dissipative quadratic soliton states. In Fig. S3a we first plot the MIC, which was obtained from Section III, to illustrate the range of modulationally unstable modes when $(\Delta_1, \beta_2) = (-5,3)$. This corresponds to the same conditions in which Fig. 3c of the main text is calculated. Fig. S3b, which is an enlarged version of Fig. 3c of the main text, illustrates that the $A^{\pm}$ solutions always exist unstably in the region of $\rho < \rho_a$ leading to both solutions being unstable throughout the entire region of the $A^0$ stable region. The parameters used in this stability analysis are provided in Section XII.

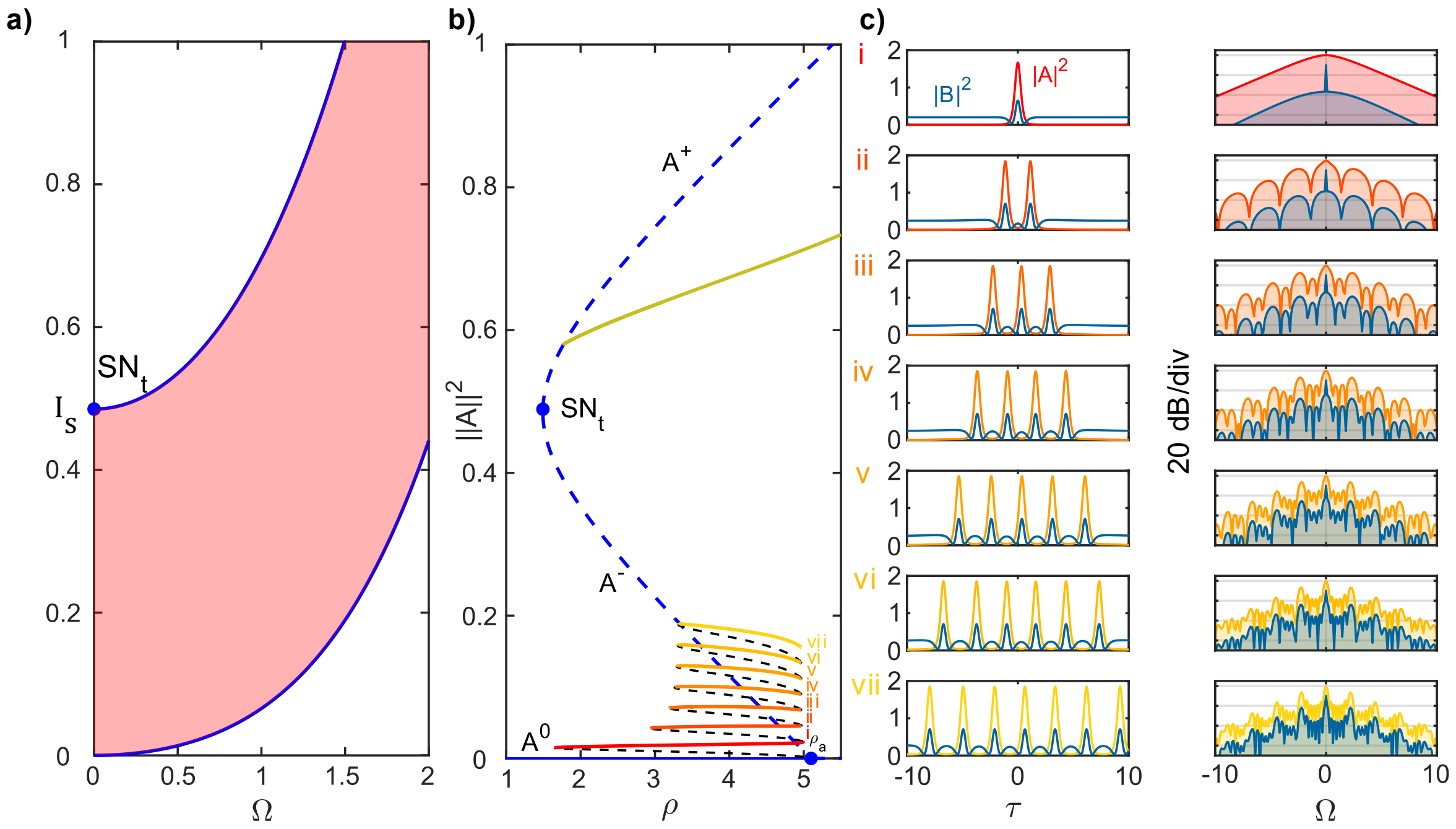

**Fig. S3: Stability analysis and Stable solutions of the DROPO system:** a) Marginal instability curve for $(\Delta_1, \beta_2, \beta_1) = (-5,3,1)$ which mirrors the condition of our DR-DOPO. Homogeneous solutions, which lie within the red area, are unstable. Within the $(||A||^2, \rho)$ – parameter space, b) illustrates that $A^{\pm}$ is always unstable (dashed-blue lines). This region corresponds to the Turing bistability regime where $A^0$ (blue-solid) coexist with the periodic temporal solution, PP (dark yellow-solid). c) Plots the temporal (left) and frequency domain (right) solutions which illustrate the first seven DQS solutions both in the signal (red) and pump (cyan-blue).

To explores this bistability region we employ path continuation methods to numerically study steady state solutions of Eq. (S23) **Error! Reference source not found.**. This is done by parameterizing around $\rho$ and applying a pseudo-arclength predictor and Newton-Raphson corrector. The calculated normalized energy of the state found during the numerical path continuation is calculated by integrating over the finite domain,

$$||A||^2 = \frac{1}{T}\int_{-\frac{T}{2}}^{\frac{T}{2}} |A|^2 \, d\tau,$$

where we set our domain to $T = 60$ (same as the main text). Fig. S3b illustrates the solution curves in the $(||A||^2 - \rho)$-parameter space which are the different stable and unstable energies of the system that can exist at each specific pump power.

We first employ the path continuation method to follow the non-trivial solution curve where we find that the dark-yellow line bifurcates subcritically from the homogenous solutions, $A^{\pm}$, corresponding to the solution for the periodic pattern (PP). The temporal solution of the PP is illustrated in the inset of Fig. 3c of the main text. The coexistence of this PP and $A^0$ is what defines the Turing bistability regime.

In this regime we then seek to find the energies and solutions which correspond to the DQS states. To do so, we again applied our numerical path-continuation method to explore the numerical solutions within the bistable region. We start the path continuation method around the subcritical pitchfork bifurcation, $\rho_a$, and initialize the field with an approximate weakly nonlinear asymptotic solution which was found in in ref [9] for an equation of the same form as Eq. $(S23)$:

$$A_w(\tau) = \left(\frac{\Delta_1}{1-\rho_a} + i\right)\frac{\sqrt{2(\rho_a-\rho)}}{C_3^2}\ \mathrm{sech}\left(\tau\frac{\sqrt{-2(\rho_a-\rho)}}{C_1^2}\right)$$

where the pulse width and amplitude are defined by

$$C_1 = -\frac{2\beta_1\Theta}{1+\Theta^2},\ \ C_3 = 2\Theta\gamma_{eff} - (\Theta^2-1), \text{with } \Theta = \frac{\Delta_1}{1-\rho_a}.$$

The path-continuation method results in a solution curve for the DQS states shown in Fig. S3b where the energy of the first seven DQS states (red to yellow solid lines) are plotted. In Fig. S3c, each DQS solution is illustrated in the temporal and frequency domain corresponding to the states i-vii. As you follow the energy states upwards, increasing in $||A||^2$, the solution states differ by a constant separation reflecting the successive addition of a soliton within the local structure. As you follow the states horizontally, increasing in $\rho$, the soliton eventually undergoes a saddle node bifurcation which links the current state with the next successive state and thus a nucleation of another soliton. This is depicted as the dashed black lines in Fig. S3b connecting the solid red through yellow lines. The cyan-blue curves in Fig. S3c illustrate the corresponding solutions for the pump field which we find through Eq. $(S24)$**.** This type of homoclinic snaking has been studied thoroughly in ref. [9,10].

Experimentally, the emergence of these solutions form spontaneously from a chaotic background during the laser scan. In the next section we study the transient cavity dynamics into this bistability regime by numerically simulating through a laser scan sweep.

## V. Numerical integration using SSFM

In this section we further probe the transient dynamics of the OPO into the Turing bistable regime. While the stability analysis conducted on Eq. $(S23)$ provides insight into the static picture of the cavity at a fixed detuning, the rescaling of the field from Eq. $(S16)$, to derive Eq. $(S23)$ does not provide an intuitive dynamical picture when the detuning evolves. This is important when conducting an analysis on the evolution of the cavity through a laser frequency sweep. Therefore, in this section, we employ a standard split step method to solve the coupled mean field equations (Eq. $(S13)$ and $(S14)$) which provide an insight into the transient dynamics of the cavity as the resonant detuning of the signal evolves into mode-locking from the red into the blue detuning. This procedure is numerically identical to the experimental procedure conducted in Fig. 3 and Fig. 4 of the main text. All the simulation parameters are provided in section XII.

Fig S4a and Fig. S4b illustrates the signal and pump intracavity energy for different cavity trajectories by integrating over the domain of the simulation ($T = 60$) while starting from different initial noise conditions. The single DQS trajectory is illustrated by the red and cyan-blue lines while other trajectories from two to seven solitons are stochastically accessed by repeating the cavity sweep with different initial noise conditions. These sweeps are illustrated as the yellow and cyan lines. These soliton solutions are identical to the solutions found by the stability analysis of Fig. S3 and are related by the rescaling value $|a|^2 = |A|^2\alpha(1+(2\Delta_1/\alpha)^2)$.

The single DQS trajectory is illustrated in the time domain in Fig. S4c and Fig. S4d and their corresponding frequency domain is illustrated in Fig. S4e and Fig. S4f. The black dashed lines illustrate transitions from different cavity solutions corresponding to the generation of two different Turing pattern states (**i-ii**) and then a transition into the chaotic intracavity state (**iii**). Finally, in this scenario, a single DQS spontaneously generates in zone **iv** from the chaotic background. This single DQS is illustrated at $\Delta_1 = -5$ in the time domain, Fig. S4g, and frequency domain, Fig. S4h. For different trajectories, the chaotic background can spontaneously form into any of the stable DQS states detailed in Fig. S3. Supplementary Video 1 animates the temporal and spectral cavity trajectory for one, two, and three cavity soliton solutions.

This transition from the chaotic state stochastically into the different DQS states exactly mirrors our experimental procedure of accessing the different soliton states plotted in the main text.

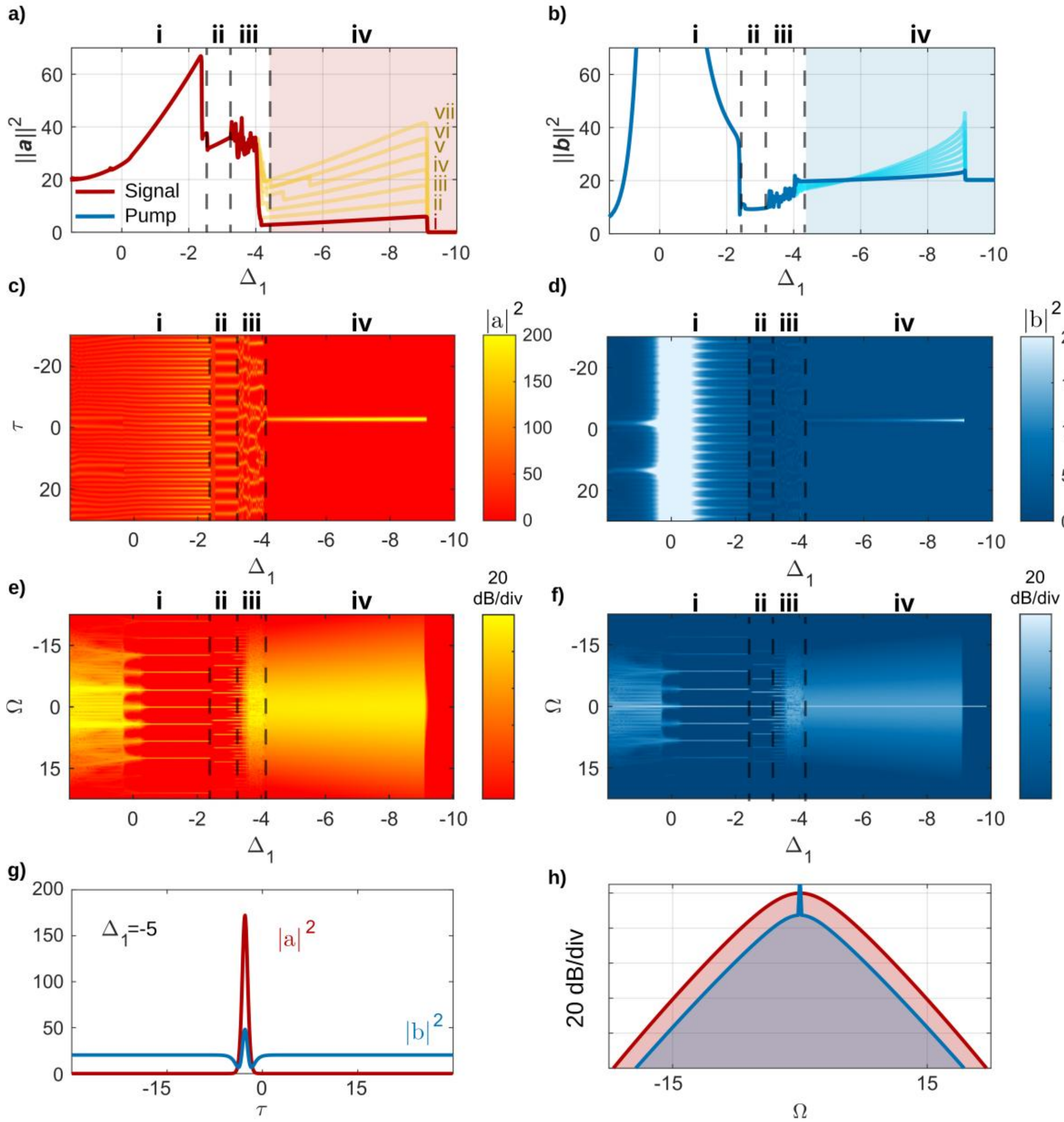

**Fig. S4 Numerical Simulations of dissipative quadratic soliton formation:** a,b) Intracavity power of the signal (red) and pump (cyan-blue) during a simulated laser scan of $2^{10}$ simulated modes over a resonance of the DR-DOPO, solving the coupled mean-field equations Eq. ($S13$)and Eq. ($S14$). The step features indicate transition from the cavity instability into the DQS state. The yellow and cyan lines trace out possible evolutions of the system during the scan indicating the formation of (i-vii) solitons. The red and blue shaded area corresponds to where solitons exist. c,d) Illustrate the temporal evolution of the single DQS trajectory and e,f) the spectral evolution. The transition between different states are labeled **i-iv** where **i** and **ii** correspond two different Turing pattern states and **iii** is the cavity chaotic state where **iv** is the transition to the low-noise DQS states. g,h) is the zoom into the soliton at $\Delta_1 = -5$. Note that the amplitudes of the coupled mean field equations and the rescaled equation are related such that $|a|^2 = \alpha(1 + 4\Delta_1^2/\alpha^2)|A|^2$and likewise for the pump. The simulation parameters are contained in Section XII.

## VI. DQS mode-locking in anomalous dispersion

According to the form of the nonlocal nonlinearity (Section I) and the corresponding stability analysis (Section III) the bistable regime identified in Eq. $(S23)$ can exist in either dispersion regime. In this manuscript we mainly are concerned with the normal dispersion regime, $\beta_1 = 1$ and set the cavity detuning to satisfy $\beta_1\Delta_1 < 0$. This condition can therefore also be reached in the anomalous regime, $\beta_1 = -1$, in the red detuning region (See Table S2).

Fig. S5a and Fig. S5b illustrates the MIC and cavity solutions in the ($||A||^2$-$\rho$)- parameter space when both the pump and signal are synthetically adjusted to be anomalous dispersion (i.e. $\beta_1 = -1$ and $\beta_2 = -3$). The homogenous and soliton solutions illustrated in Fig. S5b and Fig. S5c are found from conducting the same path continuation method outlined in Section III and IV and are identical to that of the normal dispersion regime. Fig. S5c illustrates the first three solutions in time (left) and frequency (right).

Next, we probed the transient dynamics of the anomalous dispersion regime by numerically solving Eq. $(S13)$ and Eq. $(S14)$. In this new dispersion regime, we now sweep from negative (blue) into positive (red) detuning. These transient mode-locking dynamics are exactly equivalent to the normal dispersion regime (Section V) with the exception of the stability regime now existing in the red-detuning rather than blue-detuning. This emphasizes the versatility of the DQS-modelocking to be applied to either dispersion regime and could be readily applied to compensate for perturbative effects such as thermal-refractive noise in microresonators.

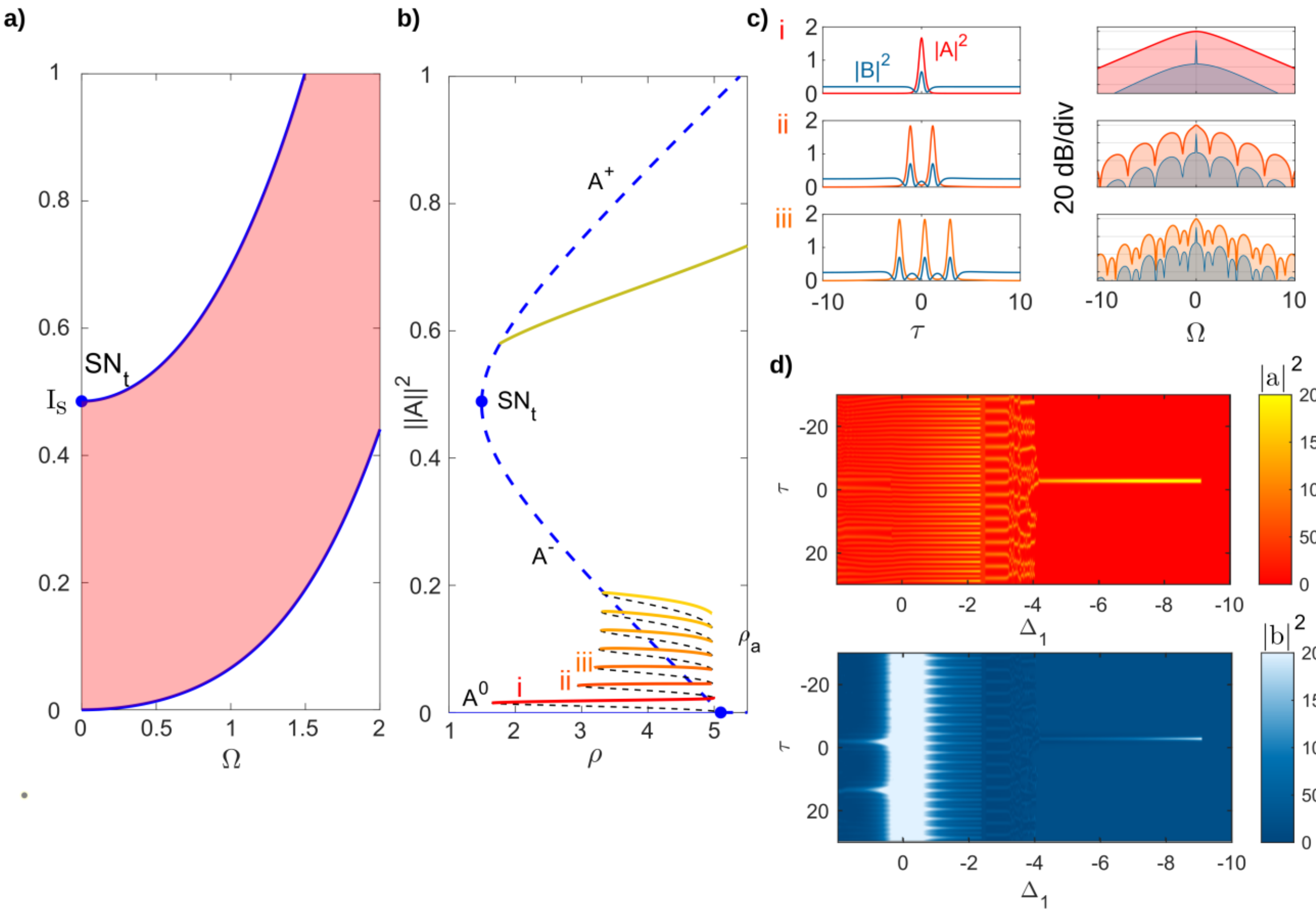

**Fig. S5: Bright DQS formation in anomalous dispersion cavity:** a) Marginal instability curve for $(\Delta_1, \beta_2, \beta_1) = (5, -3, -1)$ which mirrors the condition of our DR-DOPO but in the opposite detuning and opposite dispersion regime. Homogeneous solutions, which lie within the red area, are unstable. Within the $(||A||^2, \rho)$ – parameter space, b) illustrates that $A^{\pm}$ is always unstable (dashed-blue lines). Identically to Fig. S3b, this region corresponds to the Turing bistability regime where $A^0$ (blue-solid) coexist with the periodic temporal solution, PP (dark yellow-solid). c) Plots the temporal (left) and frequency domain (right) solutions which illustrate the first 3 DQS solutions both in the signal (red) and pump (cyan-blue). d) The single DQS cavity trajectory solving the coupled mean field equations Eq. $(S13)$ and Eq. $(S14)$. The simulation parameters are provided in Section XII. Note that c) and d) are related by the rescaling value such that $|a|^2 = \alpha(1 + 4\Delta_1^2/\alpha^2)|A|^2$ and likewise for the pump.

## VII. Phase noise measurement of DQS

In this section we further provide the phase noise of the single DQS state. The phase noise up to an offset frequency of 1 MHz is provided in Fig. S6

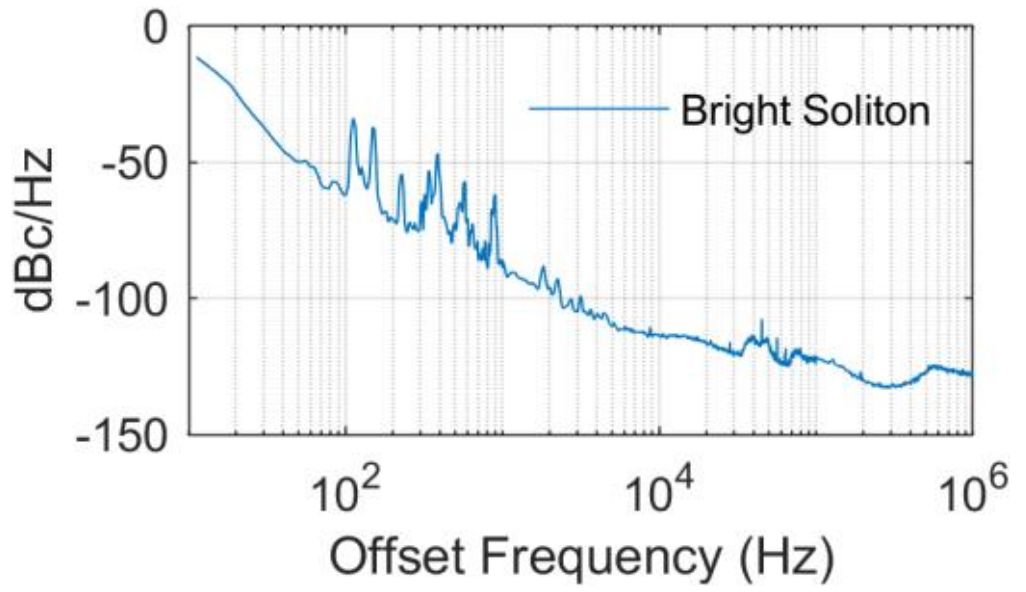

**Fig. S6 DQS Phase Noise:** The phase noise of a single DQS (Fig. 3c, State V of the main text) measured at a carrier frequency of 380 MHz.

## VIII. SHG-FROG experimental details

Fig. S7 illustrates the experimental setup to measure the spectrogram of the DQS using a second-harmonic generation (SHG) FROG. The measured spectrogram is compared to the recovered spectrogram using a recovery algorithm based on generalized projections. The recovered spectrogram has a mean-squared error of 3%.

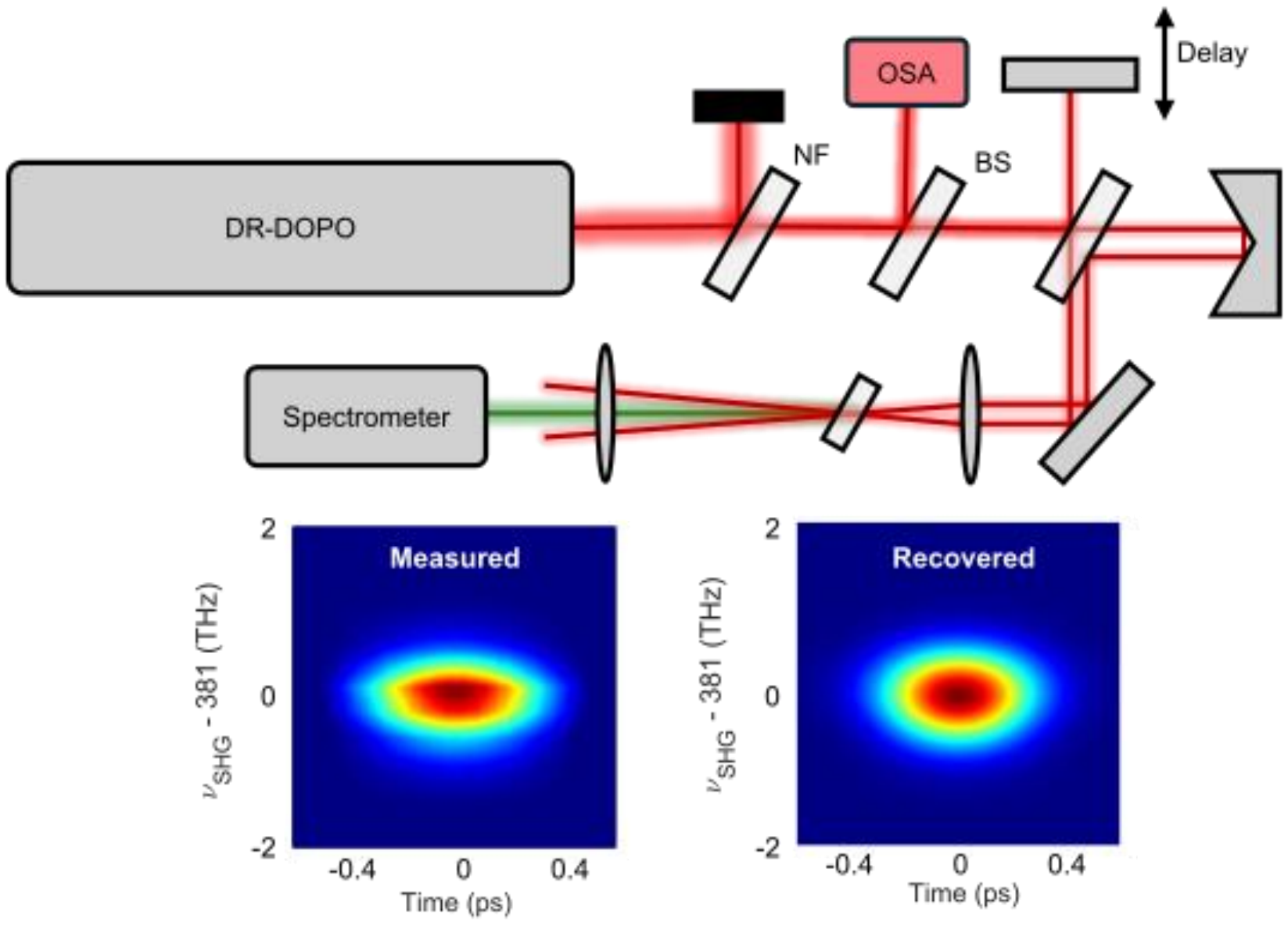

**Fig. S7 SHG-FROG expeirmental setup and measured vs recovered spectrogram.**

## IX. Effect of group velocity dispersion on bright DQS

In this section we theoretically explore the effects of GDD on soliton formation. GDD can be viewed as a second order perturbation to the nonlinear response where $Y(\Omega) = (\Delta_2 - \beta_2\Omega^2)(1 + (\Delta_2 - \beta_2\Omega^2)^2)^{-1}$, when the walk-off is set to zero. This results in an increasing bandwidth with decreasing $\beta_2$. Fig. S8a plots the EKN and ETPA as $\beta_2$ is varied at a specific detuning of $\Delta_1 = -5$ while maintaining the signal dispersion regime, $\beta_1 = 1$. As $\beta_2$ is brought from the positive into the negative, the nonlinear bandwidth quadratically increases near the zero-dispersion regime until the pump dispersion regime flips, $\beta_2 < 0$, and the narrowband phase-anomalies reappear prohibiting DQS in this regime. The near zero positive regime is ideal for engineering the broadest nonlinear response and corresponding soliton spectra. In Fig. S8b we employed the path continuation method onto Eq. (S23) on three different pump dispersion conditions illustrating the resulting DQS solutions at $\Delta_1 = -5$. The broadest possible bandwidths in both the pump and signal occur at $\beta_2 = 0$. Note that $\beta_2$ is proportional to the *ratio* of the pump to the signal dispersion thus the nonlinear response can be engineered by either the signal or pump dispersion. In the ideal case, to maximize the spectral bandwidth, both dispersions should be brought to near zero and in the same dispersion regime (i.e. either both normal or both anomalous).

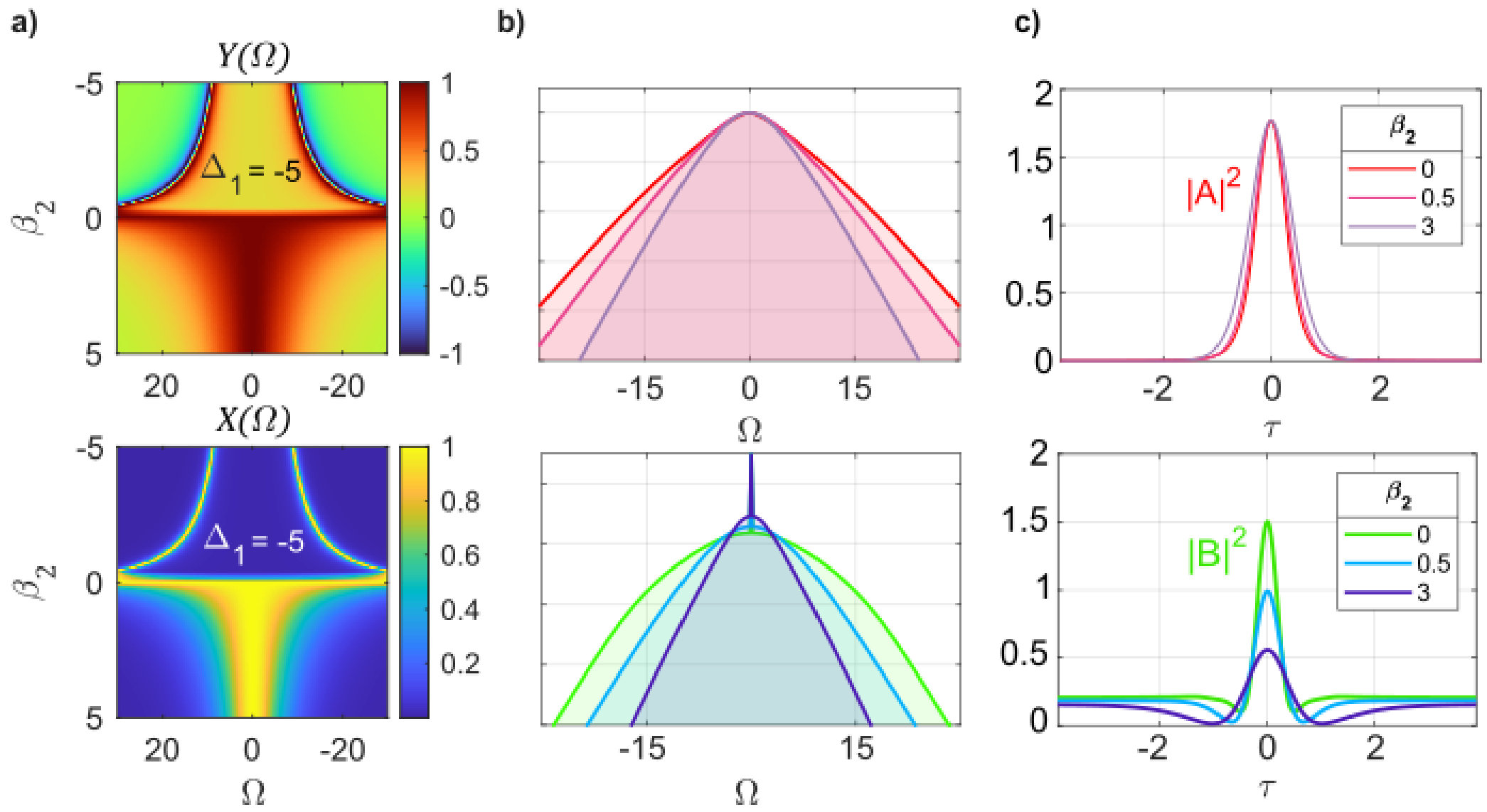

**Fig. S8: Effects of GDD on soliton formation** a) The evolution of the nonlocal nonlinearity as the dispersion ratio, $\beta_2$, is varied. The EKN (top) and ETPA (bottom) both quadratically increase in bandwidth as the dispersion is minimized near positive zero. Beyond zero, where $\beta_2 < 0$, lower zone DQS is prohibited. b,c) plots the DQS solutions for different $\beta_2$ in the spectral and temporal domain for the signal (top) and pump (bottom). The other simulation parameters are contained in Table S1.

Finally, to illustrate the influence of signal dispersion on the generated soliton, Fig. S9 presents the soliton solutions at $\Delta_1 = -5$ in physical units for varying signal dispersion, assuming a perfect pump dispersion compensation ($\beta_2 = 0$). The pulse duration scales as $(|k_1''|L)^{-1/2}$ thus the duration can be vastly improved with signal dispersion compensating techniques.

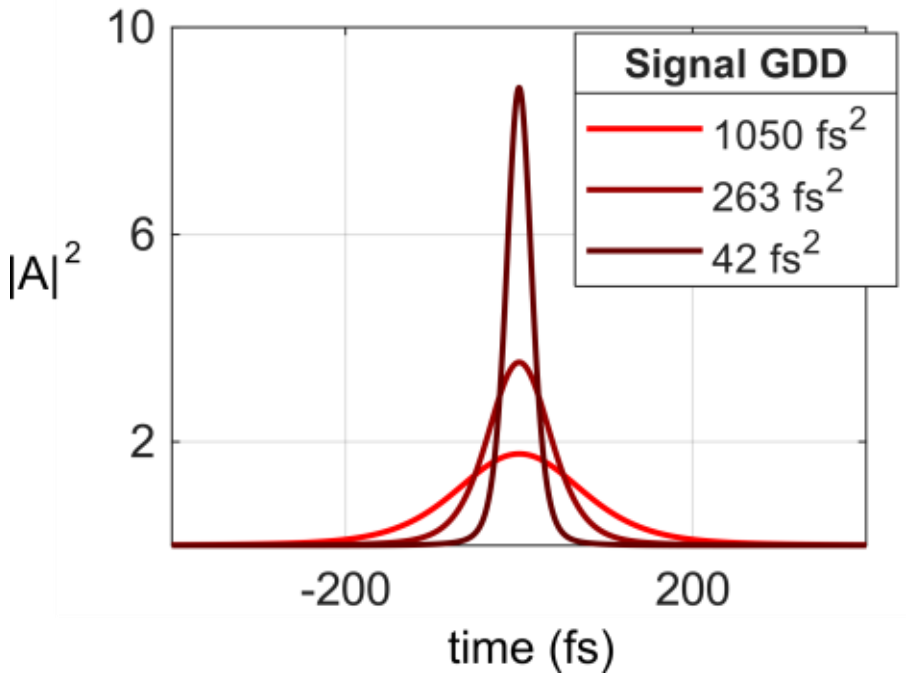

**Fig. S9:** DQS solutions at $\Delta_1 = -5$ for varying signal dispersion, assuming a perfect pump dispersion compensation ($\beta_2 = 0$).

## X. Effect of group velocity mismatch on bright DQS

According to our analysis of the nonlocal nonlinear response, we can understand the role of the GVM as a first order perturbation on the generated DQS. In the limit where GVM and GVD are *both* zero Eq. $(S20)$ becomes $Y(\Omega) = \Delta_2(1+\Delta_2^2)^{-1}$ and the nonlinear response is a flat instantaneous response. Therefore, it is straightforward that near zero GVM is imperative to ensure that the narrowband perturbations are reduced, and a smooth nonlinear response maintains the pulse quality. Fig. S10a illustrates the nonlinear response of the EKN and ETPA when different walk-off values are considered at a specific resonant detuning of $\Delta_1 = -5$. When the walk-off is small, the nonlinear bandpass is sheared laterally from the carrier frequency. This will cause an asymmetric pulse distortion like the non-instantaneous Raman effect observed in Kerr resonator-based soliton generation. This weak perturbation will not inherently prohibit soliton formation but rather distort the spectral and temporal characteristics. Figure S10b illustrates the single DQS solution calculated using the numerical path continuation method outlined in Section III and IV when $d = \pm 3$. These solutions correspond to the walk-off perturbed DQS. Fig. S10b illustrates that even weak values of walk-off can significantly change the spectrum and temporal pulse shape.

In a similar fashion to the analysis conducted in Sections V and VI, where the transient behavior of the cavity is investigated by integrating Eq. $(S13)$and $(S14)$, Fig. S10c illustrates the transient dynamics of the temporal evolution of the cavity but with the added walk-off perturbation. The generated signal and pump are temporally trapped at a constant offset from the corresponding lab frame which causes the soliton to walk-out of the simulated frame. Supplementary Video 2 animates this transient dynamic behavior of the walk-off perturbed soliton. At large walk-off values the narrowband phase-distortion reappears which will lead to a much stronger perturbation and thus lower-zone soliton is prohibited.

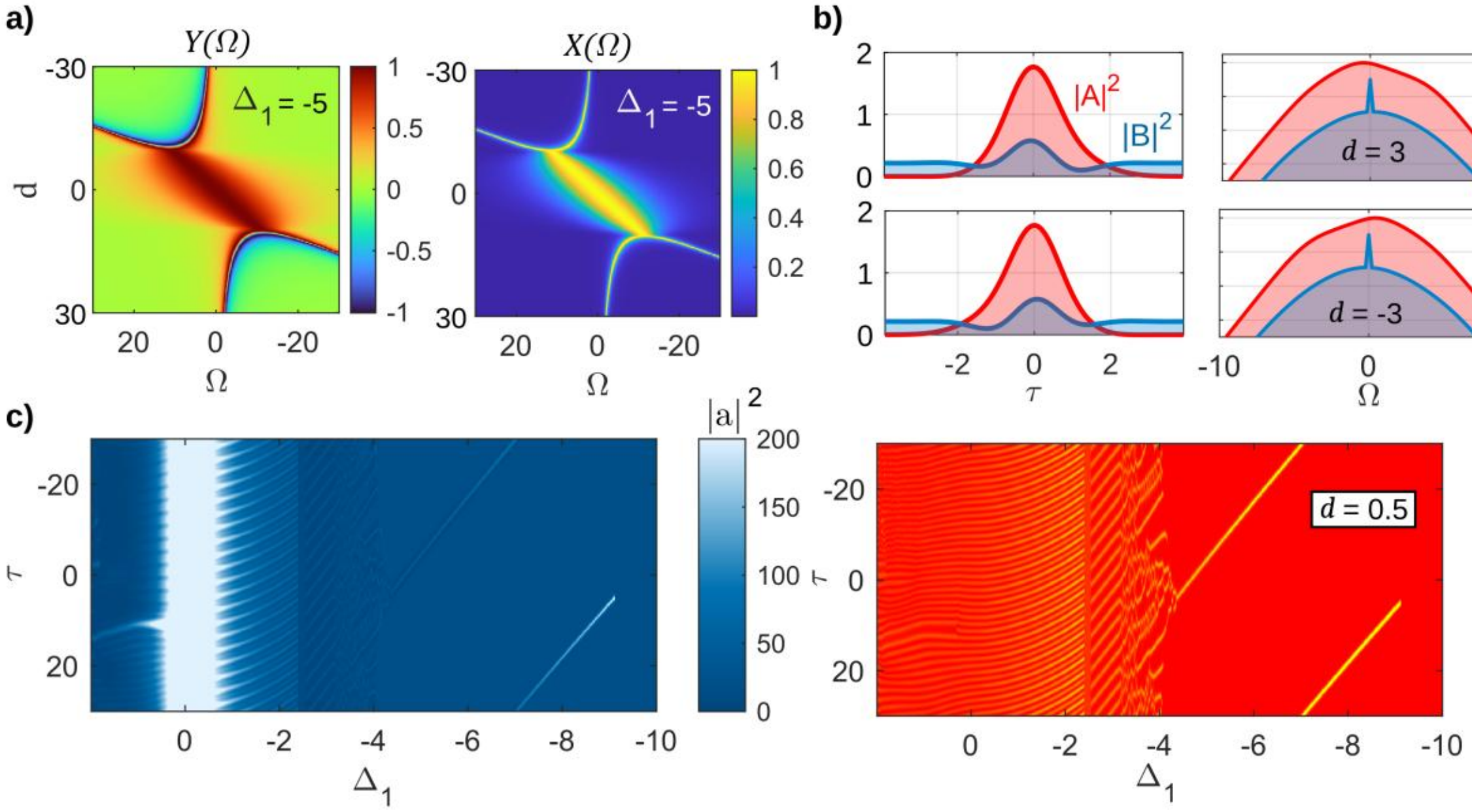

**Fig. S10: Effects of GVM on soliton formation** a) The nonlinear response at a fixed detuning $\Delta_1 = -5$ as the walk-off parameter is swept from d = $[-30,30]$. Within the smooth bandstop region, walk-off perturbed solitons may exist. Beyond this, the disruptive phase-anomalies reappear and lower zone soliton formation is prohibited. b) illustrates the walk-off perturbed soliton for $d = \pm 3$ found from the numerical path continuation method. The linear perturbed nonlinear response results in the asymmetric tilt of the temporal and spectral pulse profiles of the pump (cyan-blue) and signal (red). c) Illustrates the intracavity transient dynamics during a laser can in the presence of walk-off for the pump (left) and signal (right) for $d = 0.5$. The perturbation causes the soliton to drift out of the lab frame the other simulation parameters are contained in Section XII. Note that b) and c) are related by the rescaling parameter such that $|a|^2 = \alpha(1 + (2\Delta_1)^2/\alpha^2)|A|^2$ and likewise for the pump. Supplementary Video 2 illustrates this cavity trajectory during the laser scan.

In the experiment, the zero-GVM wavelength was calculated to be ~786 nm at a PPLN crystal temperature of 28.7°C. In Fig. S11, we experimentally achieved the walk-off perturbated soliton by tuning the pump wavelength away from the zero-GVM point. Fig. S11a illustrates the same laser scan as Fig. 3 of the main text, where the pump is now set to a wavelength of 785nm. The dynamics are nearly identical to that of the zero-GVM case, however each transient spectra are observed to have a slight asymmetric tilt associated with the walk-off perturbation of the nonlinear response. Line profiles of this laser scan are provided in Fig. S11b which illustrate the primary and secondary comb generation $(I, II)$ and into the modulationally unstable states $(III, IV)$. With further detuning, the cavity spontaneously collapses into $V$, the single walk-off DQS solution. The black-dashed lines illustrate the theoretical cavity spectra while Fig. S11e illustrates the numerical cavity solution in the time domain corresponding to a walk-off value of $d = 5$.

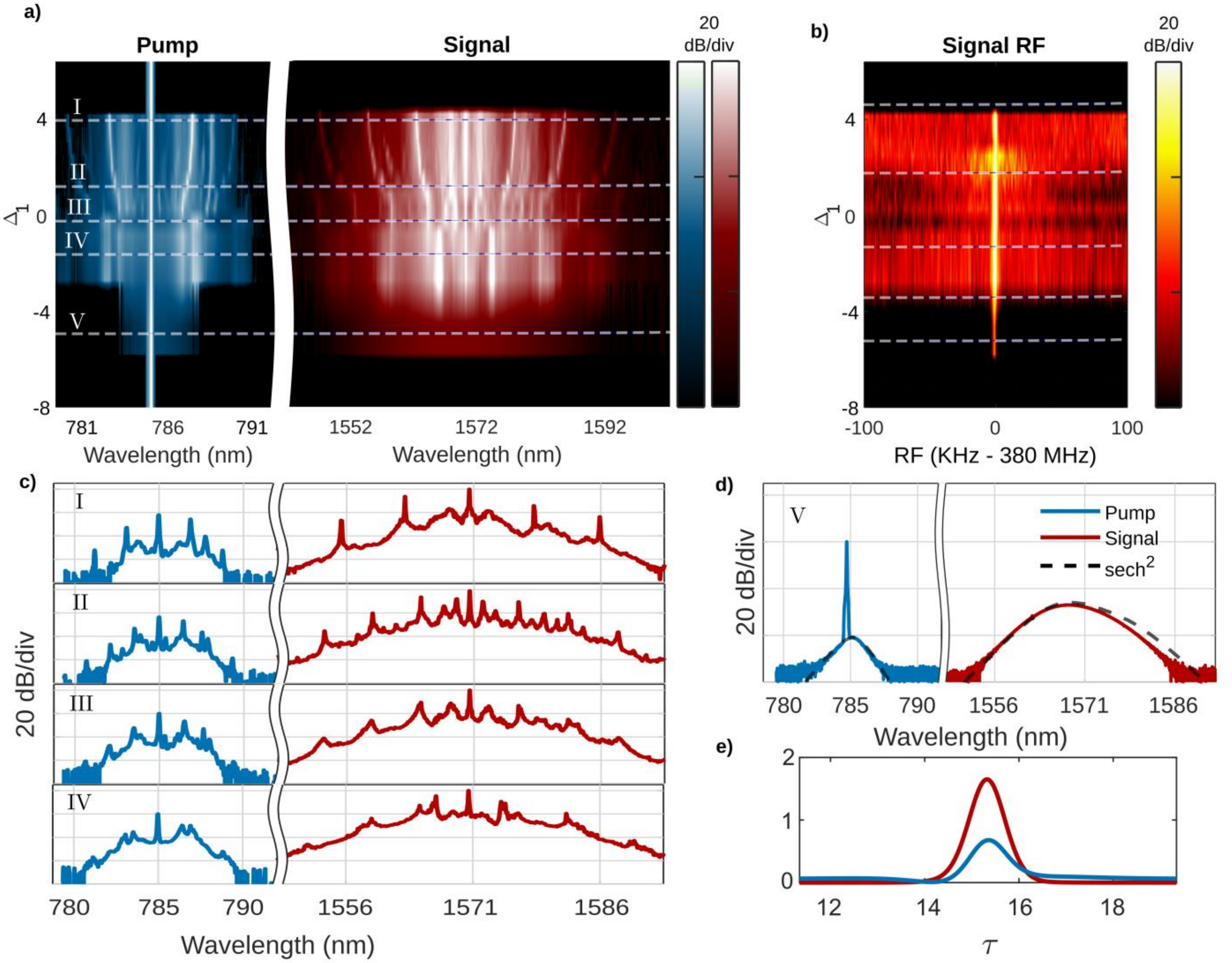

**Fig. S11: Experimental demonstration of Walk-Off perturbed DQS mode-locking** a) Experimental spectra captured for the pump (cyan-blue) and signal (red) is simultaneously measured as the 640 mW CW pump laser is brought into the resonance. Concurrently, b) captures the signal radio frequency (RF) spectra measured about the repetition rate of the DR-DOPO at an RBW of 1KHz. As the pump is brought into the resonance, c) illustrates line-profiles of the captured pump and signal spectra for points along the laser scan. These states correspond to the walk-off perturbed analog of Fig. 3c of the main text. Here the state I) is the primary and II) the secondary comb formation which transitions from these coherent states into modulationally unstable state III) and finally chaotic intracavity field IV). d) illustrates the final walk-off perturbed soliton for the pump and signal where we plot the numerically generated spectra corresponding to $d = 5$. e) Plots the temporal profile.

Compared to a singly resonant system, the DR-DOPO is much more sensitive to the overlap of the interacting field. Therefore, while DR-DOPO can greatly reduce the soliton threshold, the design requirements of the cavity GVM are more stringent compared to singly resonant systems [13,14]. In the doubly resonant system, a simple solution to avoid the walk-off effects would be to move to a deeper red detuning or to increase the nonlinear response bandwidth by engineering the cavity dispersion. According to Fig. S1c a deeper red detuning comes at the cost of a decreased EKN and higher pump threshold.

In the experiment, the zero-GVM wavelength is limited by the intersection of the temperature dependent GVM and the available poling period of our PPLN which enables the quasi-phase matching. In the type-1 interaction this intersection is found at a wavelength of 786nm and a temperature of 28.7°C. In general, the GVM zero condition is an important concern in designing the DQS DR-DOPO system and many mature crystal platforms currently exist which exhibit a natural zero-GVM at a degenerate frequency. Some common materials are provided in Fig. S12a. These natural GVM zero points conveniently span the NIR to MIR almost completely and indicate the DQS principle can be deployed across a wide range of platforms using convenient quasi-phase matching techniques such as periodic poling (PP) or orientation patterning (OP). Besides, the natural GVM zero point can be conveniently temperature tuned, observed by Fig. S12b. However, since the phase mismatch condition strongly depends on crystal temperature the poling period will need to be adjusted accordingly to achieve the original phase-matched condition.

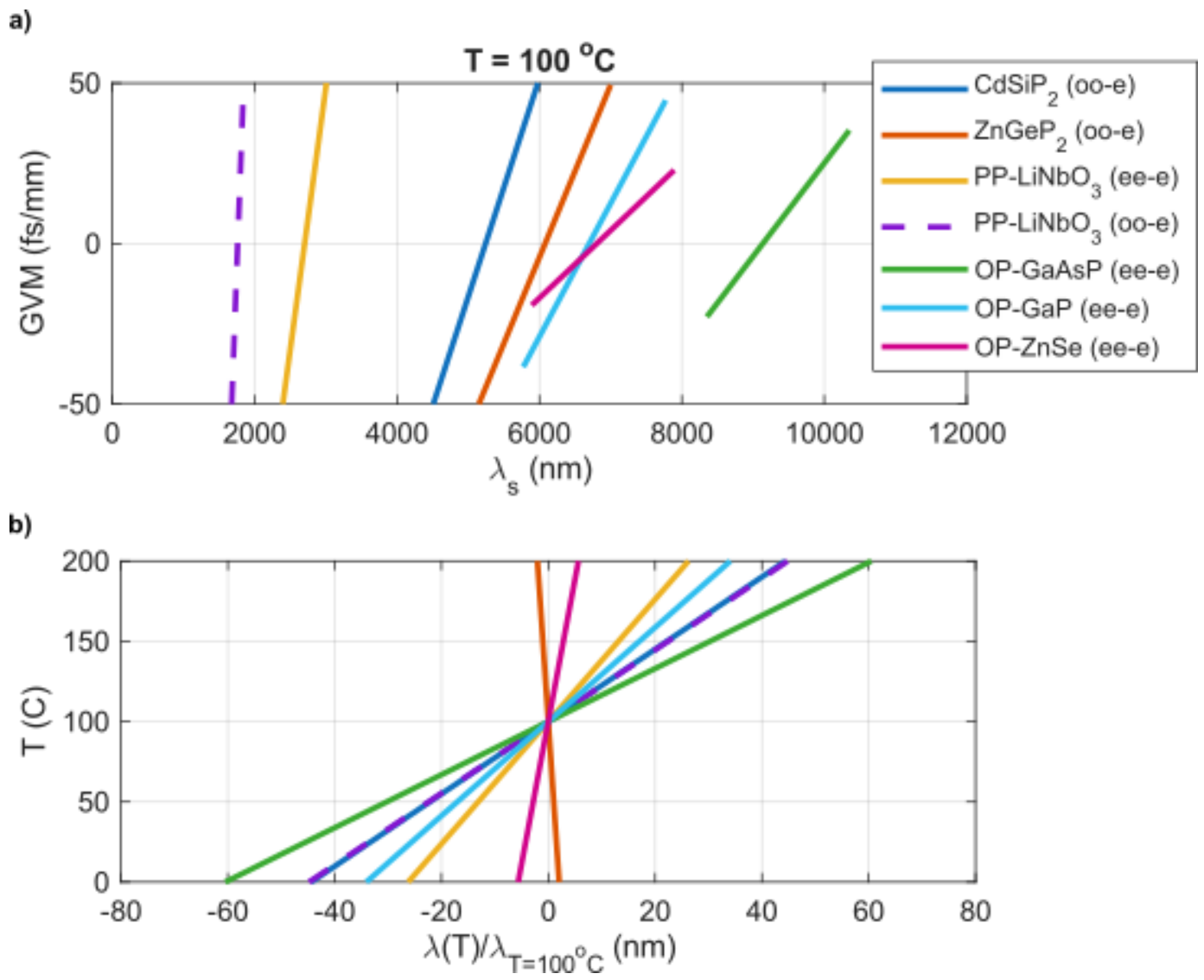

**Fig. S12:** a) GVM as a function of the degenerate signal wavelength for different $\chi^{(2)}$ materials and b) the corresponding zero-GVM point of the respective crystals as a function of crystal temperature. Type-1 phase-matched degenerate operation in PP-$LiNbO_3$ (dashed line) is used in the DR-DOPO of this manuscript.

## XI. Conversion efficiency of the DQS

In this section, we examine the achievable conversion efficiency of the DQS mode-locked OPO. Neglecting ETPA and assuming an instantaneous nonlinear response, Eq. (S23) further simplifies to:

$$\partial_\tau \mathrm{A} = -(1 + i\Delta_1)\mathrm{A} - i\beta_1 \partial_\tau^2 \mathrm{A} + i\Delta_2 \mathrm{A}\, |\mathrm{A}|^2 + \rho \mathrm{A}^* \tag{S40}$$

whose solutions obey:

$$0 = -(1 + i\Delta_1)\mathrm{A} - i\beta_1 \partial_\tau^2 \mathrm{A} + i\Delta_2 \mathrm{A}\, |\mathrm{A}|^2 + \rho \mathrm{A}^* \tag{S41}$$

Although these assumptions are less accurate than the numerical continuation method used in Section III, they admit analytical solutions that provide insight into the fundamental limitations of DQS mode-locked OPO. Substituting the ansatz $\mathrm{A}(\tau) = u(\tau)e^{i\phi}$ into Eq. (S23) and separately equating the real and imaginary parts yields two sech-shaped solutions,

$$u_\pm = P_\pm \operatorname{sech}\left(\frac{\tau}{T}\right).$$

with the phase satisfying

$$\cos(2\phi) = \rho^{-1}$$

Here, $P_\pm = \sqrt{2\Delta_{eff}^{\pm}/\Delta_2}$ and $T = \sqrt{2\beta_1/P_\pm^2\Delta_2}$ with the effective detuning defined as $\Delta_{eff}^{\pm} = \Delta_1 \pm \sqrt{\rho^2 - 1}$. Previous analysis has shown that only the higher-peak-power branch is stable against perturbations, whereas the lower-peak-power branch is a saddle point and is therefore unstable [11]. Applying the soliton existence condition, $\beta_1\Delta_1 < 0$, with $\beta_1 = 1$ in our experiment, identifies $u_-$ as the stable solution and its pulse energy is

$$E = \frac{4}{-\Delta_2}\sqrt{\beta_1\left(\Delta_1 - \sqrt{\rho^2 - 1}\right)}.$$

An important subtlety is that the normalization used in Eq. (S23) is defined for a specific value of $\Delta_2$. Consequently, $\rho$ depends on $\Delta_2$ even when the physical pump power, $S^2$, is held fixed. To eliminate this normalization dependence, we express the energy below in units of $|\mathrm{a}|^2$ rather than $|\mathrm{A}|^2$, such that the calculation of conversion efficiency, $\eta$, is more straightforward.

Figure S13a compares the analytical soliton solution with numerical results that include ETPA. In the absence of pump dispersion, the nonlinear response is instantaneous and the analytical solution agrees closely with the full numerical solution, with the remaining discrepancy arising primarily from ETPA. Unlike conventional TPA, ETPA does not constitute a true dissipative loss; instead, it enhances back-conversion from the signal to the pump near the soliton peak. Figure S13b shows the signal-soliton energy and corresponding conversion efficiency as functions of the maximum detuning, $\Delta_1^{\mathrm{Max}}$, which scales with the normalized pump amplitude, $S$, in a manner similar to that in DKS systems [12]. However, the conversion efficiency predicted in this dispersion-free limit is substantially lower than that observed experimentally.

This discrepancy can be understood by including pump dispersion in the numerical model. Figure S13c shows the soliton peak power and duration as functions of the pump-to-signal dispersion ratio, $\beta_2$. Non-zero pump dispersion renders the nonlinear response non-local, causing the analytical and numerical solutions to diverge. It also limits the soliton bandwidth and hence the minimum transform-limited pulse duration, as discussed in Section IX. By contrast, the carrier-frequency magnitudes of the EKN and ETPA remain unchanged, leaving the peak power comparatively insensitive to pump dispersion. Overall, the conversion efficiency increases with pump dispersion, as shown in Fig. S13d.

Figures S13c,d thus show that the instantaneous-response analytical model does not fully describe the DR-DOPO in the presence of pump dispersion. Instead, it establishes lower bounds on the conversion efficiency and transform-limited pulse duration. Although increasing the pump dispersion can distort the pulse and raise the soliton threshold, it also provides an additional degree of freedom for controlling the DQS solution. The pump and signal dispersions are therefore intrinsically coupled to the non-local nonlinear response and can be co-engineered to optimize different operating regimes.

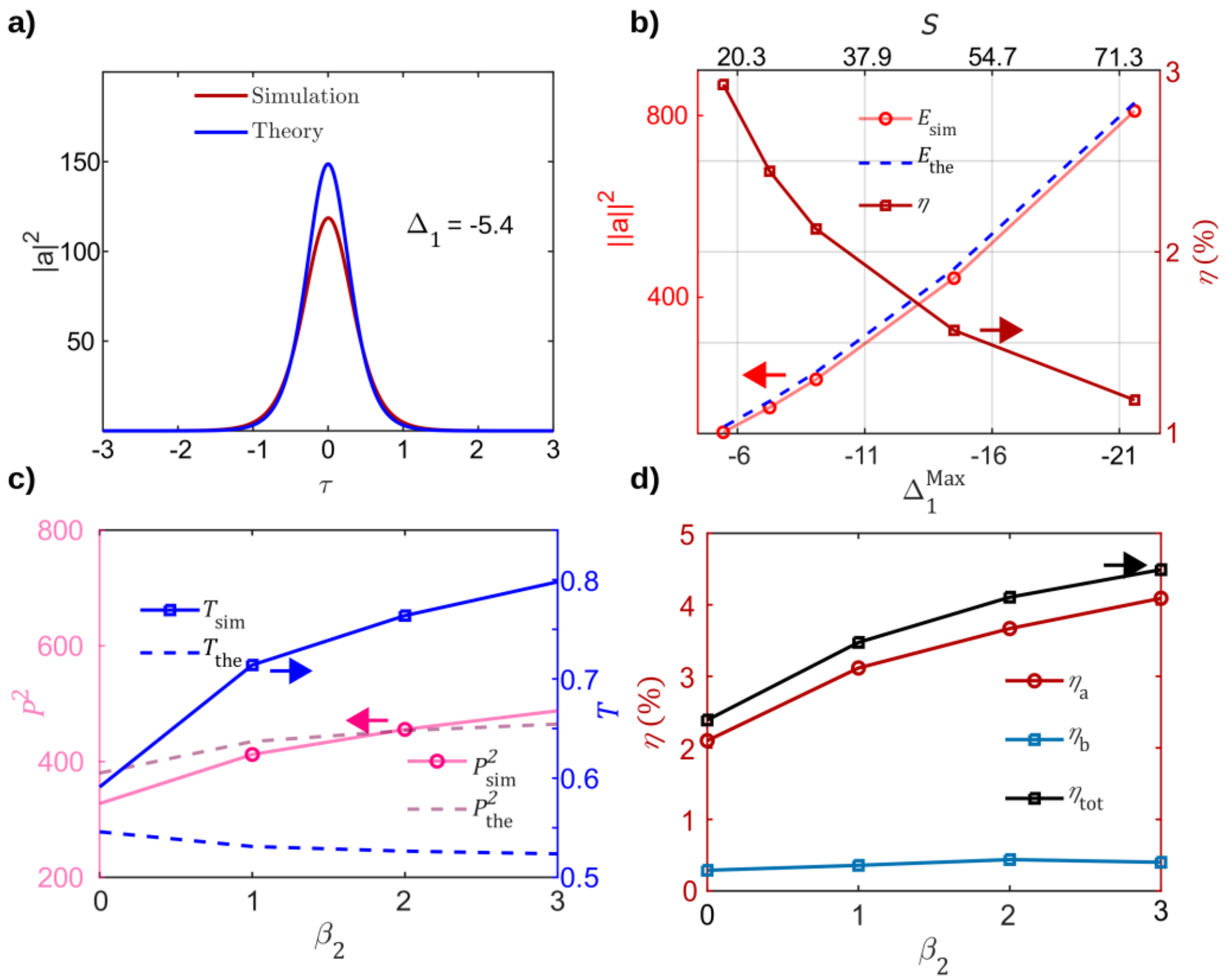

**Fig. S13: Soliton Conversion Efficiency** a) the theoretical analytic soliton solution (blue) neglecting the effects of ETPA compared to the numerically integrated solution (red) including the ETPA term where the nonlinear response is instantaneous $\beta_2 = 0$. b) The simulated energy (left) compared to the theoretical as a function of the pump amplitude ($S$) calculated at the edge of the soliton existence detuning range ($\Delta_1^{Max}$). The conversion efficiency of the signal soliton is plotted on the right axis at each soliton energy c) the effects of the pump dispersion on the soliton solution for the peak power (pink) and soliton duration (blue) as a function of the pump dispersion. d) the conversion efficiency, $\eta_a, \eta_b$, and $\eta_{tot}$, corresponding to the signal, pump, and total efficiency respectively, wherein the black arrow corresponds to the parameters that most closely resemble our experimental condition.

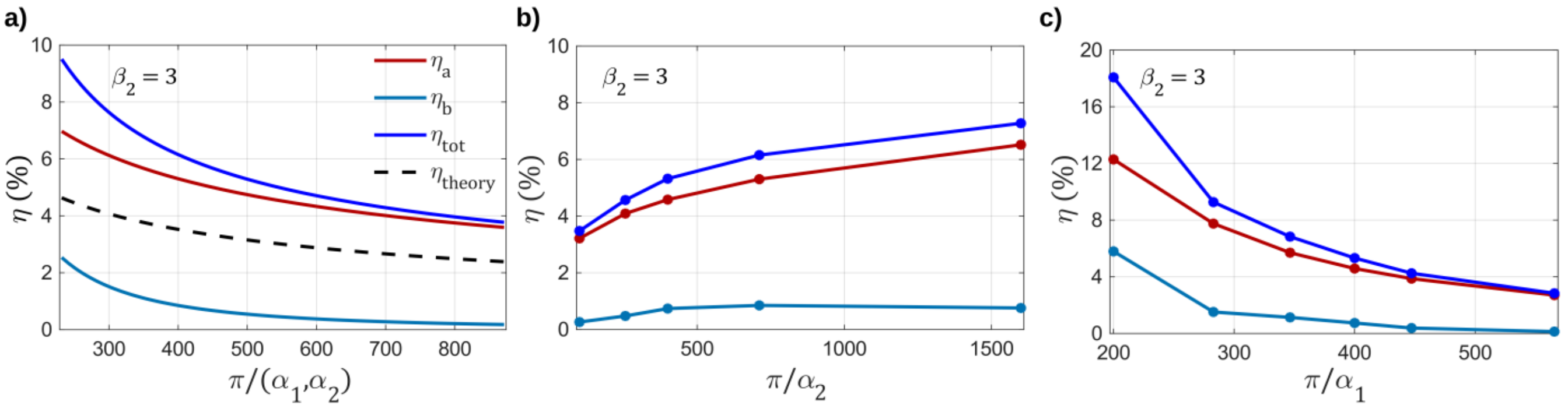

**Fig. S14: Conversion efficiency dependence on pump and signal linear loss:** a) Simulated conversion efficiency at $\beta_2 = 3$ of the signal (red), pump (cyan-blue), and combined (blue) compared to the theoretical (black) which assumes no ETPA nor nonlinear dispersion as both the pump and signal cavity finesse are varied equally while maintaining critical coupling. b) The simulated conversion efficiency as the signal maintains a finesse of $\pi/400$ and the pump finesse is varied, both pump and signal maintain critical coupling. c) Simulated conversion efficiency as the pump maintains a finesse of $\pi/400$ and the signal finesse is varied, both pump and signal maintain critical coupling.

Figure S14 illustrates how the pump-to-DQS conversion efficiencies depend on linear cavity loss. In Fig. S14a, the pump and signal finesses are varied simultaneously while both fields remain critically coupled. Increasing the cavity finesse reduces the ETPA perturbation and suppresses back-conversion from the signal DQS (red) to the pump DQS (cyan-blue), thereby monotonically decreasing the pump DQS contribution to the total DQS conversion efficiency (blue). Conversely, a lower finesse increases cavity extraction, enhancing the DQS conversion efficiencies of both fields.

Figure S14b shows the corresponding results when the signal finesse is fixed at $\pi/400$ and the pump finesse is varied under critical coupling. Increasing the pump finesse enhances the intracavity pump field, thereby lowering the pump threshold and increasing the DQS conversion efficiency. Conversely, Fig. S14c fixes the pump finesse at $\pi/400$ while varying the signal finesse. In this case, the pump enhancement remains approximately constant, whereas reducing the signal finesse improves signal extraction. Consequently, the DQS conversion efficiency increases monotonically as the signal approaches the non-resonant limit, albeit at the expense of a higher pump threshold.

## XII. Experimental details

In this section we provide further details into the experimental demonstration of DQS-mode locking. Fig. S15a plots a thorough map of the pumping and stabilization procedure. Fig. S15b plots a MZI calibrated measurement of the cavity linewidth. Finally, Fig. S15c plots the theoretical change in the phase match (blue) compared to the change in the FSR between pump and signal (red) as a function of the temperature. Fig. S15c is included to illustrate the nearly orthogonalc control of the doubly resonant condition using the translation stage (TS) compared to the phase matching using the PPLN temperature.

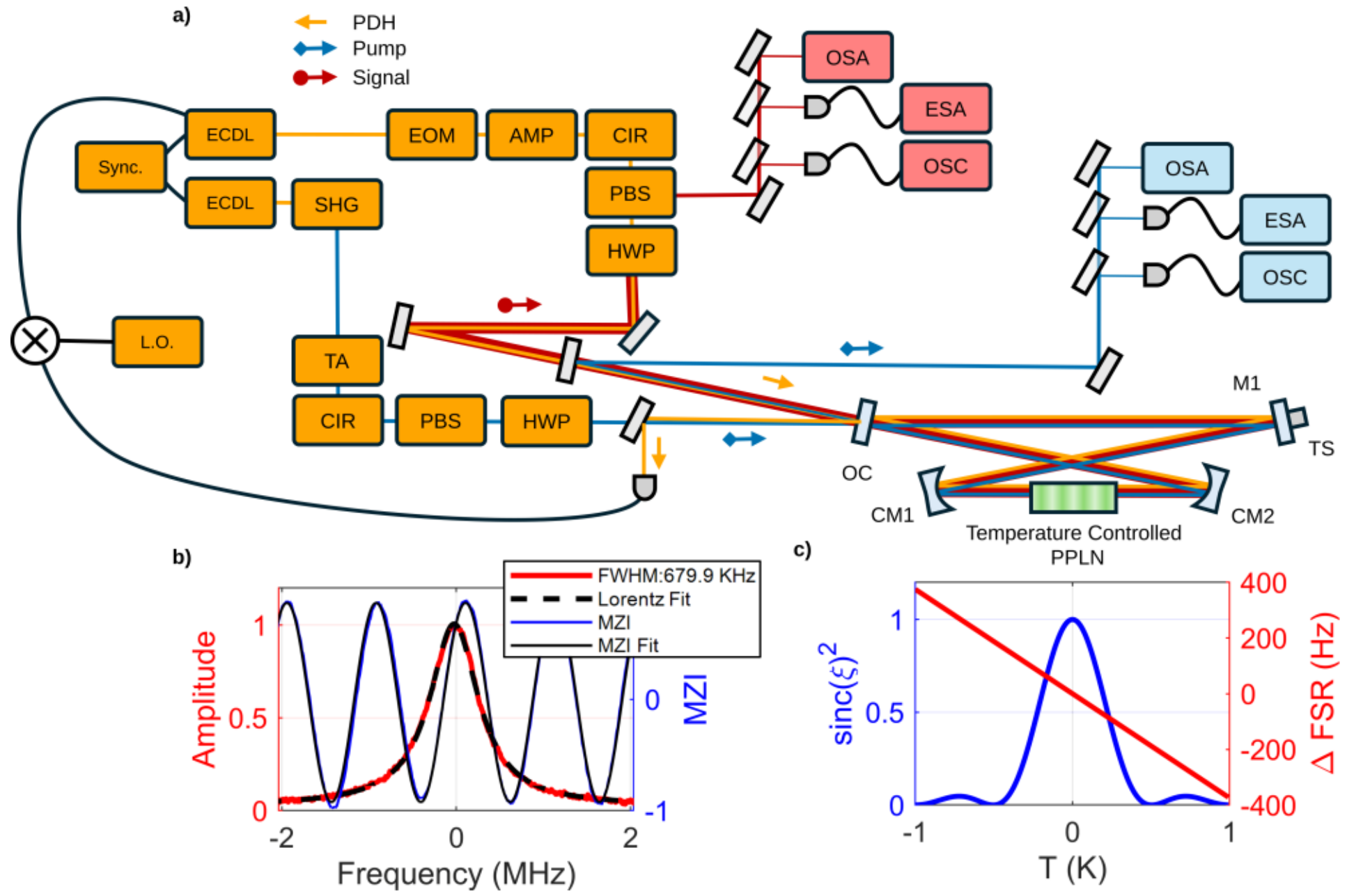

**Fig. S15: Experimental setup** a) The two ECDLs are phase-locked to each other the top ECDL is used for PDH error signal (orange) and bottom ECDL is used for the pump (cyan-blue) independently b) cavity finesse measured by a calibrated MZI. c) the theoretical change in the phase-matching (blue) compared to the difference between pump and signal FSR ($\Delta FSR$) as a function of Temperature. External cavity diode laser (ECDL), electro-optic modulator (EOM), Erbium doped fiber amplifier (AMP), circulator (CIR), polarizing beam splitter (PBS), half-wave plate (HWP), oscilloscope (OSC), electrical spectrum analyzer (ESA), optical spectrum analyzer (OSA), control electronics for PDH (C.E.), comb-line band pass filter (CL-BPF), output coupler (OC), flat mirror (M), translation stage (TS), curved mirror (CM).

## XIII. Lists of acronyms and physical parameters

| Acronym | Name |
|---|---|
| **MLL** | Mode-locked lasers |
| **OPO** | Optical parametric oscillator |
| **DQS** | Dissipative quadratic soliton |
| **CW** | Continuous wave |
| **DR-DOPO** | Doubly resonant degenerate optical parametric oscillator |
| **PICQN** | Phase-matched intracavity cascaded quadratic nonlinearity |
| **EKN** | Effective Kerr nonlinearity |
| **ETPA** | Effective two photon absorption |
| **PPLN** | Periodically poled lithium niobate |
| **SR** | Singly resonant |
| **DR** | Doubly resonant |
| **TPA** | Two photon absorption |
| **MKN** | Material Kerr nonlinearity |
| **GVD** | Group velocity dispersion |
| **GVM** | Group velocity mismatch |
| **PP** | Periodic patterns |
| **RF** | Radio frequency |
| **RBW** | Radio bandwidth |
| **FSR** | Free spectral range |
| **DKS** | Dissipative Kerr soliton |
| **FROG** | Frequency resolved optical gating |

**Table S3: List of acronyms used in the main text**

| Parameter | Value | Units | Definition | Name |
|---|---|---|---|---|
| $\boldsymbol{\omega_0}$ | 120 | $2\pi$ THz | | Carrier frequency |
| $\boldsymbol{d_{eff}}$ | 2.7 | pm/V | | Effective nonlinear coefficient |
| $\boldsymbol{n_p}$ | 2.208353 | unitless | | Index of refraction, pump |
| $\boldsymbol{n_s}$ | 2.177959 | unitless | | Index of refraction, signal |
| $\boldsymbol{L}$ | 10 | mm | | Cavity length |
| $\boldsymbol{\tilde{\tau}}$ | | s | | Fast time, physical units |
| $\boldsymbol{\tau}$ | | unitless | $\tilde{\tau}(2\alpha_1/\lvert k_1''\rvert L)^{1/2}$ | Normalized fast time |
| $\boldsymbol{\nu}$ | | Hz | | Frequency, physical units |
| $\boldsymbol{\Omega}$ | | unitless | $\nu/(2\alpha_1/\lvert k_1''\rvert L)^{1/2}$ | Normalized frequency |
| $\boldsymbol{\tilde{z}}$ | | m | | Propagation position, physical units |
| $\boldsymbol{\tilde{t}}$ | | s | | Slow time, physical units |
| $\boldsymbol{t}$ | | unitless | $\tilde{t}(\alpha_1/t_r)$ | Normalized slow time |
| $\boldsymbol{m}$ | | Int. | | Round trip index |
| $\boldsymbol{\Delta T}$ | | unitless | | Normalized pulse duration |
| $\boldsymbol{t_r}$ | 75 | ps | | Round trip time |
| $\boldsymbol{T}$ | 60 | unitless | | Normalized round trip time |
| $\boldsymbol{\Delta k}$ | 0 | 1/m | | Phase mismatch |
| $\boldsymbol{\xi}$ | 0 | unitless | $\Delta kL/2$ | Normalized phase mismatch |
| $\boldsymbol{\psi}$ | | unitless | $\xi/2 + \tan^{-1}(-\Delta_2/\alpha)/2$ | Rescaled phase mismatch |
| $\boldsymbol{\Delta k'}$ | 0 | s/m | | Group velocity mismatch |
| $\boldsymbol{d}$ | 0 | Arb. | $\Delta k'(2L/\alpha_1\lvert k_1''\rvert)^{1/2}$ | Normalized group velocity mismatch |
| $\boldsymbol{k_1''}$ | 108.015 | $fs^2$/mm | | Group velocity dispersion, signal |
| $\boldsymbol{k_2''}$ | 388.958 | $fs^2$/mm | | Group velocity dispersion, pump |
| $\boldsymbol{\beta_1}$ | 1 | unitless | $sign(k_1'')$ | Signal dispersion regime |
| $\boldsymbol{\beta_2}$ | 3 | unitless | $\lvert k_2''\rvert/k_1''$ | Pump-to-signal dispersion ratio |
| $\boldsymbol{A_{eff}}$ | 5026.5 | $\mu m^2$ | | Effective mode area |
| $\boldsymbol{\kappa}$ | 0.13 | 1/(Wm) | | Second-order nonlinear coupling coefficient |
| $\boldsymbol{\alpha_{c1,2}}$ | 0.8 | 1/m | | Loss per unit length |
| $\widetilde{\mathbf{E_p}}, \widetilde{\mathbf{E_s}}$ | | $W^{1/2}$ | | Electric field envelopes |
| $\mathbf{a}, \mathbf{b}$ | | $1/W^{1/2}$ | $(E_s, E_p)\kappa L/\alpha_1$ | Normalized field envelopes |
| A,B | | $1/W^{1/2}$ | $a, b/e^{i\psi}\sqrt{\alpha(1+(\Delta_2/\alpha)^2)}$ | Rescaled field envelopes |
| $\mathbf{I_s}$ | | 1/W | $\lvert A\rvert^2$ | Rescaled signal power |
| $\boldsymbol{E_{in}}$ | $\sqrt{600}$ | $W^{1/2}$ | | Input electric field |
| $\boldsymbol{S}$ | 45 | unitless | $\sqrt{\theta_2}B_{in}\kappa L/\alpha_1^2$ | Normalized pump power |
| $\boldsymbol{\rho}$ | 4.5 | unitless | $S/\alpha\sqrt{1+(\Delta_2/\alpha)^2}$ | Rescaled input pump amplitude |
| $\boldsymbol{\theta_{1,2}}$ | 0.008 | unitless | | Coupler transmission |
| $\boldsymbol{\delta_1}$ | -0.04 | rads | | Signal phase detuning |
| $\boldsymbol{\delta_2}$ | -0.08 | rads. | | Pump phase detuning |
| $\boldsymbol{\alpha_{1,2}}$ | 0.008 | unitless | | Total cavity loss per round trip |
| $\boldsymbol{\Delta_1}$ | $-5$ | unitless | $\delta_1/\alpha_1$ | Normalized phase detuning, signal |
| $\boldsymbol{\Delta_2}$ | $-10$ | unitless | $\delta_2/\alpha_1$ | Normalized phase detuning, pump |
| $\boldsymbol{\alpha}$ | 1 | unitless | $\alpha_2/\alpha_1$ | Pump to signal round trip loss ratio |
| $\boldsymbol{\delta_0(\Omega)}$ | | unitless | $\delta_0(0) = 1,\ \delta_0(\Omega \neq 0) = 0$ | Kronecker Delta function |
| $\boldsymbol{\hat{J}(\Omega)}$ | | unitless | $\dfrac{\text{sinc}^2(\xi)}{\alpha + i(\Delta_2 - d\Omega - \beta_2\Omega^2)}$ | Frequency domain nonlinear response |
| $\boldsymbol{J(\tau)}$ | | unitless | $J(\tau) = \mathcal{F}\{J(\Omega)\}$ | Temporal nonlinear response |
| $\boldsymbol{\bar{J}(\tau)}$ | | unitless | $\alpha(1+(\Delta_2/\alpha)^2)J(\tau)$ | Rescaled nonlinear response |
| $\boldsymbol{X(\Omega)}$ | | unitless | $Re\{\hat{J}(\Omega)\}$ | Dispersive effective two photon absorption |
| $\boldsymbol{Y(\Omega)}$ | | unitless | $Im\{\hat{J}(\Omega)\}$ | Dispersive effective Kerr nonlinearity |
| $\boldsymbol{\alpha_{ETPA}}$ | | | $\kappa^2LX(0) = \kappa^2Lsinc^2(\xi)\dfrac{\alpha\alpha_1}{(\alpha\alpha_1)^2 + (\Delta_2\alpha_1)^2}$ | Effective two photon absorption |
| $\boldsymbol{\gamma_{EKN}}$ | | | $\kappa^2LY(0) = \kappa^2Lsinc^2(\xi)\dfrac{\Delta_2\alpha_1}{(\alpha\alpha_1)^2 + (\Delta_2\alpha_1)^2}$ | Effective Kerr nonlinearity |

**Table S4: List of physical, normalized, and rescaled parameters used in the experiments and simulations**